\documentclass[utf8,a4]{frontiersFPHY} 
\setcitestyle{square} 
\usepackage{url,lineno,microtype,subcaption}
\usepackage[onehalfspacing]{setspace}
\usepackage{amssymb,amsmath,amsthm,amsfonts,amscd,amsbsy}
\newtheorem{theorem}{Theorem}
\usepackage{booktabs}
\usepackage{multirow}
\usepackage[colorlinks=true, allcolors=blue, breaklinks=true]{hyperref}
\newcommand{\sech}{\mathrm{sech} \,}

\def\keyFont{\fontsize{8}{11}\helveticabold }
\def\firstAuthorLast{N. Karjanto} 
\def\Authors{N. Karjanto\,$^{*}$}
\def\Address{Department of Mathematics, University College, Natural Science Campus, Sungkyunkwan University, 2066 Seobu-ro, Suwon, 16419, Republic of Korea}
\def\corrAuthor{Corresponding Author}
\def\corrEmail{natanael@skku.edu}

\begin{document}
\onecolumn
\firstpage{1}
\title[Peregrine soliton]{Peregrine soliton as a limiting behavior of the Kuznetsov-Ma and Akhmediev breathers}
\author[\firstAuthorLast ]{\Authors} 
\address{} 
\correspondance{} 
\extraAuth{}%
\maketitle

\begin{abstract}
This article discusses a limiting behavior of breather solutions of the focusing nonlinear Schr\"odinger (NLS) equation. These breathers belong to the families of solitons on a non-vanishing and constant background, where the continuous-wave envelope serves as a pedestal. The rational Peregrine soliton acts as a limiting behavior of the other two breather solitons, i.e., the Kuznetsov-Ma breather and Akhmediev soliton. Albeit with a phase shift, the latter becomes a nonlinear extension of the homoclinic orbit waveform corresponding to an unstable mode in the modulational instability phenomenon. All breathers are prototypes for rogue waves in nonlinear and dispersive media. We present a rigorous proof using the $\epsilon$-$\delta$ argument and show the corresponding visualization for this limiting behavior.

\tiny
\keyFont{\section{Keywords:} nonlinear Schr\"odinger equation, Kuznetsov-Ma breather, Akhmediev soliton, Peregrine soliton, waves on a non-vanishing and constant background, limiting behavior, modulational instability, rogue waves}
\end{abstract}

\section{Introduction}

The Peregrine soliton, also known as the rational solution, is one of the solutions of the focusing nonlinear Schr\"odinger (NLS) equation. Analyzed and derived for the first time by Peregrine in 1983, its characteristic is localized in both space and time~\citep{peregrine1983water}. The Peregrine soliton was successfully observed experimentally in nonlinear optics~\citep{kibler2010peregrine}, water waves~\citep{chabchoub2011rogue}, and multi-component plasma~\citep{bailung2011observation}. Together with the Kuznetsov-Ma breather and Akhmediev soliton, the Peregrine soliton belongs to the families of soliton solutions of the NLS equation on a non-vanishing but the constant background, where the plane-wave or continuous-wave solution acts for such a pedestal.

The NLS equation is a nonlinear evolution equation that models slowly varying envelope dynamics of a weakly nonlinear quasi-monochromatic wave packet in dispersive media. The model has an infinite set of conservation laws and belongs to a completely integrable system of nonlinear partial differential equations through the Inverse Scattering Transform (IST). It has a wide range of applications in various physical settings, such as surface water waves, nonlinear optics, plasma physics, superconductivity, and Bose-Einstein condensates (BEC).

The NLS equation in nonlinear optics was first derived by Kelley in 1965 using a nonlinear electromagnetic wave Maxwell's equation introduced by Chiao et al. one year earlier~\citep{kelley1965self,chiao1964self}. Furthermore, Karpman and Krushkal in 1969 derived the NLS equation using the Whitham-Lighthill adiabatic approximation, where the original article in Russian was published one year earlier~\citep{karpman1969modulated}. Tappert and Varma also derived the NLS equation for heat pulses in solids using the asymptotic theory in 1970~\citep{tappert1970asymptotic}. 

In 1968, Taniuti and Washimi derived the NLS equation describing dispersive hydromagnetic waves propagating along an applied magnetic field in a cold quasi-neutral plasma using the method of multiple scales~\citep{taniuti1968self}. To the best of our knowledge, it is this article that mentions, introduces, and hence popularizes the term ``nonlinear Schr\"odinger equation'' for the first time. One year later, Taniuti, Yajima, and Asano derived the NLS equation using the perturbation method in an electron plasma wave system that admits plane-wave solution with high-frequency oscillation and the nonlinear Klein-Gordon equation~\citep{taniuti1969perturbation,asano1969perturbation}.

The NLS equation was first derived independently in hydrodynamics by Benney and Newell in 1967 for wave packet envelopes propagation and the long time behavior of weakly interacting waves~\citep{benney1967propagation} and in 1968 by Zakharov for deep-water waves using a spectral method~\citep{zakharov1968stability}. The NLS equation for gravity water waves with uniform and finite depth was derived by Hasimoto and Ono in 1972 using singular perturbation methods~\citep{hasimoto1972nonlinear}.

In the field of BEC, the NLS equation with the non-zero potential term is known by another name: the Gross-Pitaevskii equation, where both Gross and Pitaevskii independently derived this equation in 1961~\citep{gross1961structure,pitaevskii1961vortex}. In the field of superconductivity, the time-independent NLS equation resembles some similarities with a simplified $(1+1)$-D form of the Ginzburg-Landau equation derived a decade earlier in 1950~\citep{ginzburg2009theory}. For a further overview of the NLS equation, see the encyclopedia articles~\citep{malomed2005nlse,ablowitz2008nlssystem} and book chapters~\citep{huang2018nlse,karjanto2020nlse}. For an extensive discussion of the NLS equation, see the monographs~\citep{sulem2007nonlinear,fibich2015nonlinear}.

The NLS equation admits exact analytical solutions, and some families of these solutions, known as ``breather solitons'', are excellent prototypes for rogue wave modeling in various nonlinear media. An explicit expression of the breather solitons will be presented in Section~\ref{exactsec}. In what follows, we provide an overview of breather solitons and their connection with rogue wave phenomena.

There are various excellent reviews on rogue wave phenomena based on the NLS equation as a mathematical model and its corresponding breather solitons. Onorato et al. covered rogue waves in several physical contexts including surface gravity waves, photonic crystal fibers, laser fiber systems, and 2D spatiotemporal systems~\cite{onorato2013brogue}. Dudley et al. reviewed breathers and rogue waves in optical fiber systems with an emphasis on the underlying physical processes that drive the appearance of extreme optical structures~\cite{dudley2014instabilities}. They reasoned that the mechanisms driving rogue wave behavior depend very much on the system. Residori et al. presented physical concepts and mathematical tools for rogue wave description~\cite{residori2017rogue}. They highlighted the most common features of the phenomenon include large deviations of wave amplitude from the Gaussian statistics and large-scale symmetry breaking. Chen et al. discussed rogue waves in scalar, vector, and multidimensional systems~\cite{chen2017versatile} while Malomed and Mihalache surveyed some theoretical and experimental studies on nonlinear waves in optical and matter-wave media~\cite{malomed2019nonlinear}.

Rogue waves come from and are closely related to modulational instability with resonance perturbation on continuous background~\cite{zhao2016quantitative}. A comparison of breather solutions of the NLS equation with emergent peaks in noise-seeded modulational instability indicated that the latter clustered closely around the analytical predictions~\cite{toenger2015emergent}. ``Superregular breathers'' is the term coined indicating creation and annihilation dynamics of modulational instability, and the evidence of the broadest group of these superregular breathers in hydrodynamics and optics has been reported~\cite{kibler2015superregular}. An interaction between breather and higher-order rogue waves in a nonlinear optical fiber is characterized by a trajectory of localized troughs and crests~\cite{liu2018interaction}.

Breather soliton solutions find several applications, among others in beam-plasma interactions~\cite{veldes2013electromagnetic}, in the transmission line analog of a nonlinear left-handed metamaterials~\cite{shen2017solitons}, in a nonlinear model describing an electron moving along the axis of a deformable helical molecules~\cite{albares2018solitons}, and in the mechanisms underlying the formation of and real-time prediction of extreme events~\cite{farazmand2019extreme}. Additionally, optical rogue waves also successfully simulated in the presence of nonlinear self-image phenomenon in the near-field diffraction of plane waves from light wave grating, known as the Talbot effect~\cite{zhang2014nonlinear}.

Since the definitions of ``rogue waves'' and ``extreme events'' are varied, a roadmap for unifying different perspectives could stimulate further discussion~\cite{akhmediev2016roadmap}. Theoretical, numerical, and experimental evidence of the dissipation effect on phase-shifted Fermi-Pasta-Ulam-Tsingou (FPUT) dynamics in a super wave tank, which is related to modulational instability, can be described by the breather solutions of the NLS equation~\cite{kimmoun2016modulation}. Since the behavior of a large class of perturbations characterized by a continuous spectrum is described by the identical asymptotic state, it turns out that the asymptotic stage of modulational instability is universal~\cite{biondini2016universal}. Surprisingly, the long-time asymptotic behavior of modulationally unstable media is composed of an ensemble of classical soliton solutions of the NLS equation instead of the breather-type solutions~\cite{biondini2016oscillation}.

There are various techniques to derive the breather solutions of the focusing NLS equation, among others are the phase-amplitude algebraic ansatz~\cite{akhmediev1985generation,akhmediev1986modulation,akhmediev1987exact,akhmediev1997solitons}, the Hirota method~\cite{hirota1974new,hirota1976direct,hirota1976variety,ablowitz1990homoclinic}, nonlinear Fourier transform IST~\cite{ablowitz1981solitons,newell1985solitons,osborne2000nonlinear,osborne2001random,biondini2014inverse}, symmetry reduction methods~\cite{olver2000applications}, variational formulation and displaced phase-amplitude equations~\cite{van2006displaced,karjanto2006mathematical,karjanto2007derivation}. Another derivation using IST with asymmetric boundary conditions is given in~\cite{demontis2014inverse}.

General $N$-solitonic solutions of the NLS equation in the presence of a condensate derived using the dressing method describe the nonlinear stage of the modulational instability of the condensate~\cite{gelash2014superregular}. Recently, both theoretical description and experimental observation of the nonlinear mutual interactions between a pair of copropagative breathers are presented and it is observed that the bound state of breathers exhibit a behavior similar to a molecule with quasiperiodic oscillatory dynamics~\cite{xu2019breather}.

The purpose of this paper is to revisit the connection between the families of the breather soliton solutions, both analytically and visually. The paper will be presented as follows. After this introduction, Section~\ref{exactsec} presents several important exact solutions of the NLS equation. In particular, the focus will be on the breather type of solutions. Section~\ref{limitbehave} discusses a rigorous proof for the limiting behavior of the breather wave solutions using the $\epsilon$-$\delta$ argument. The limiting behavior will continue in Section~\ref{libevisual}, where we cover it from the visual point of view. We present the corresponding contour plots for various values of parameters and the parameterization sketches of the non-rapid oscillating complex-valued breather amplitudes. Finally, Section~\ref{conclusion} concludes our discussion and provide remarks for potential future research.

\section{Exact solutions of the NLS equation}	\label{exactsec}

Throughout this article, we adopt the following $(1+1)$-dimension, focusing-type of the NLS equation in a standard form:
\begin{equation}
i q_t + q_{xx} + 2 |q|^2 q = 0, \qquad q(x,t) \in \mathbb{C}.
\end{equation}
Usually, the variables $x$ and $t$ denote the space and time variables, respectively. The simplest-solution is called the plane-wave or continuous-wave solution: $q(x,t) = q_0(t) = e^{2it}$. Another simple solution with a vanishing background is known as the bright soliton or one-soliton solution, given as follows:
\begin{equation}
q(x,t) = q_\text{S}(x,t) = a \, \sech(a x - 2 a b t + \theta_{0}) e^{i(b x \, + \, (a^2 \,- \, b^2)t \, + \, \phi_0)}, \qquad a, b, \theta_{0}, \phi_0 \in \mathbb{R}.
\end{equation}
Zakharov and Shabat obtained this solution in the 1970s using the IST~\cite{zakharov1972exact,zakharov1973interaction}. 

Throughout this article, our discussion will be focused on the type of NLS solutions with constant and non-vanishing background, sometimes also called the families of ``breather soliton solutions''~\citep{dysthe1999note}. There are three families of breathers, and all of them are considered as weakly nonlinear prototypes for freak waves events. Other solutions of the NLS equation include cnoidal wave envelopes that can be expressed in terms of the Jacobi elliptic functions and can be derived using the Hirota bilinear transformation, theta functions, or with some clever algebraic ansatz~\cite{chow1995class,akhmediev1997solitons}.

Otherwise mentioned, our coverage will also follow the historical order of the time when the breathers are found. Thus, when both breathers are discussed, we usually cover the Kuznetsov-Ma breather before the Akhmediev soliton. Furthermore, the term ``breather'' and ``soliton'' can be used interchangeably in this article, and they can also appear as a single term ``breather soliton''. All of them refer to the same object, i.e., the exact analytical solutions of the NLS equation with a non-vanishing, constant pedestal, or background of continuous-wave solution.\\

\subsection{The Kuznetsov-Ma breather}
The first one is called the Kuznetsov-Ma breather, where Kuznetsov, Kawata and Inoue, and Ma derived it independently in the late 1970s~\citep{kuznetsov1977solitons,kawata1978inverse,ma1979perturbed}. So, perhaps a more accurate name for this breather is the KKIM breather, which stands for the ``Kuznetsov-Kawata-Inoue-Ma breather''. However, ever since the breather dynamics are observed experimentally in optical fibers in 2012~\cite{kibler2012observation}, the former name is getting more popular even though a similar term ``Kuznetsov-Ma soliton'' has been introduced earlier~\cite{slunyaev2006nonlinear}. Hence, we adopt and use the terminology ``Kuznetsov-Ma breather'' throughout this article. We denote it as $q_\text{M}$ and is explicitly given by
\begin{equation}
q(x,t) = q_\text{M}(x,t) = e^{2it} \left(\frac{\mu^3 \cos(\rho t) + i \mu \rho \sin(\rho t)}{2 \mu \cos(\rho t) - \rho \cosh(\mu x)} + 1 \right)  \label{KMbreather}
\end{equation}
where $\rho = \mu \sqrt{4 + \mu^2}$. The Kuznetsov-Ma breather does not represent a traveling wave. It is localized in the spatial variable $x$ and periodic in the temporal variable $t$, and hence some authors also called it as the ``temporal periodic breather''~\cite{grimshaw2001wave}. 

A minor typographical error found in Kawata and Inoue's paper~\cite{kawata1978inverse} has been corrected by Gagnon~\cite{gagnon1993solitons}. Kawata and Inoue~\cite{kawata1978inverse}, as well as Ma~\cite{ma1979perturbed}, derived the Kuznetsov-Ma breather solution using the IST for finite boundary conditions at $x \rightarrow \pm \infty$. The derivation using a direct method of B\"acklund transformation can be found in~\cite{adachihara1988solitary,mihalache1993two}, where the former analyzed solitary waves in the context of an optical bistable ring cavity.

Defining the amplitude amplification factor (AF) as the quotient of the maximum breather amplitude and the value of its background~\cite{karjanto2006mathematical}, we obtain that the amplitude amplification for the Kuznetsov-Ma breather is always greater than the factor of three and is explicitly given by
\begin{equation}
\text{AF}_\text{M}(\mu	) = 1 + \sqrt{4 + \mu^2}, \qquad \mu > 0.
\end{equation}
The function is bounded below and is increasing as the parameter $\mu$ also increases. The plot of this AF can be found in~\cite{karjanto2006mathematical,karjanto2007derivation}, and different expressions of AF for this breather also appear in~\cite{clamond2006long,onorato2013brogue,chabchoub2014hydrodynamics,residori2017rogue}.

The Kuznetsov-Ma breather finds applications as a rogue wave prototype in nonlinear optics~\cite{akhmediev1992phase,akhmediev1997solitons,kibler2012observation} and deep-water gravity waves~\cite{osborne2000nonlinear,osborne2001random,kharif2001focusing,chabchoub2014hydrodynamics}. A numerical comparison of the Kuznetsov-Ma breather indicated that a qualitative agreement was reached in the central part of the corresponding wave packet and on the real face of the modulation~\cite{clamond2006long}. The stability analysis of the Kuznetsov-Ma breather using a perturbation theory based on the IST verified that although the soliton is rather robust with respect to dispersive perturbations, damping terms strongly influence its dynamics~\cite{garnier2011inverse}.

Dynamics of the Kuznetsov-Ma breather in a microfabricated optomechanical array showed an excellent agreement between theory and numerical calculations~\cite{xiong2017kuznetsov}. The spectral stability analysis of this breather has been considered using the Floquet theory~\cite{cuevas2017floquet}. The mechanism of the Kuznetsov-Ma breather has been discussed and two distinctive mechanisms are paramount: modulational instability and the interference effects between the continuous-wave background and bright soliton~\cite{zhao2018mechanism}. New scenarios of rogue wave formation for artificially prepared initial conditions using the Kuznetsov-Ma and superregular breathers in small localized condensate perturbations are demonstrated numerically by solving the Zakharov-Shabat eigenvalue problem~\cite{gelash2018formation}.
 
A higher-order Kuznetsov-Ma breather can be derived using the Hirota method and utilized in studying soliton propagation with the presence of small plane-wave background~\cite{belanger1996bright,tajiri1998breather}; or using the bilinear method~\cite{chow1995solitary}.\\

\subsection{The Akhmediev soliton}
The second one is called the Akhmediev-Eleonski\u{i}-Kulagin breather and was found in the 1980s~\citep{akhmediev1985generation,akhmediev1986modulation,akhmediev1987exact}. In short, we simply call it the ``Akhmediev soliton'' and denote it as $q_\text{A}$. This breather is localized in the temporal variable $t$ and is periodic in the spatial variable $x$, and it can be written explicitly as follows:
\begin{equation}
q(x,t) = q_\text{A}(x,t) = e^{2it} \left(\frac{\nu^3 \cosh(\sigma t) + i \nu \sigma \sinh(\sigma t)}{2 \nu \cosh(\sigma t) - \sigma \cos(\nu x)} - 1 \right). 
\label{AEKbreather}
\end{equation}
Here, the parameter $\nu$, $0 \leq \nu < 2$ denotes a modulation frequency (or wavenumber) and $\sigma(\nu) = \nu \sqrt{4 - \nu^2}$ is the modulation growth rate. The colleagues from nonlinear optics prefer calling this soliton as ``instanton'' instead of ``breather'' since it breathers only once~\cite{turitsyn2012dispersion}. Other names for this solution include ``modulational instability''~\cite{akhmediev1997solitons}, ``homoclinic orbit''~\cite{ablowitz1990homoclinic,calini2002homoclinic}, ``spatial periodic breather''~\cite{grimshaw2001wave}, and ``rogue wave solution''~\cite{osborne2001random}.

The amplitude amplification for the Akhmediev soliton is at most of the factor of three and is explicitly given by
\begin{equation}
\text{AF}_\text{A}(\nu) = 1 + \sqrt{4 - \nu^2}, \qquad 0 < \nu < 2.
\end{equation}
This function is bounded above and below, $1 < \text{AF}_\text{A} < 3$, and is decreasing for an increasing value of the modulation parameter $\nu$. Although the maximum growth rate occurs for $\nu = \sqrt{2}$, the maximum AF occurs when $\nu \rightarrow 0$, when the Akhmediev breather becomes the Peregrine soliton. To the best of our knowledge, this expression was introduced by Onorato et al. in their study on freak wave generation in random ocean waves where this AF depends on the wave steepness and number of waves under the envelope~\cite{onorato2001occurrence}. The plot for this AF can be found in~\cite{karjanto2011investigation,karjanto2006mathematical,akhmediev2009waves}. Some variations in the AF expression for this soliton also appear in~\cite{clamond2006long,onorato2011triggering,onorato2013brogue,slunyaev2013highest,chabchoub2014hydrodynamics,residori2017rogue}.

The Akhmediev soliton is rather well-known due to its characteristics being a nonlinear extension of linear modulational instability. This instability is also known as sideband (or Bespalov-Talanov) instability in nonlinear optics~\cite{bespalov1966filamentary,ostrovskii1967propagation,karpman1967self}, or Benjamin-Feir instability in water waves~\cite{benjamin1967disintegration,zakharov1968stability}. Some authors studied the modulational instability in plasma physics~\cite{taniuti1968self,tam1969amplitude,hasegawa1970observation,hasegawa1972theory} and in BEC~\cite{robins2001modulational,konotop2002modulational,smerzi2002dynamical,baizakov2002regular,salasnich2003modulational,theocharis2003modulational}.  Modulational instability is defined as the temporal growth of the continuous-wave NLS solution due to a small, side-band modulation, in a monochromatic wave train. A geometric condition for wave instability in deep water waves is given in~\cite{lighthill1965contributions} and for a historical review of modulational instability, see~\cite{zakharov2009modulation}.

It has been shown numerically and experimentally that the modulated unstable wave trains grow to a maximum limit and then subside. In the spectral domain, the wave energy is transferred from the central frequency to its sidebands during the wave propagation for a certain period, and then it is recollected back to the primary frequency mode~\cite{yuen1975nonlinear,lake1977nonlinear,yuen1978relationship,yuen1978fermi,yuen1980instabilities}. It turns out that the long-time evolution of these unstable wave trains leads to a sequence of modulation and demodulation cycles, known as the FPUT recurrence phenomenon~\cite{fermi1955studies,janssen1981modulational}. Although the FPUT recurrence using the NLS model has been observed experimentally in surface gravity waves in the late 1970s~\cite{lake1977nonlinear}, it took more than two decades for the phenomenon to be successfully recovered in nonlinear optics~\cite{van2001experimental}.

Since the modulational instability extends nonlinearly to the Akhmediev soliton, it is no surprise that the former is considered as a possible mechanism for the generation of rogue waves while the latter acts as one prototype~\cite{janssen2003nonlinear,dysthe2008oceanic,kharif2008rogue}. For wave trains with amplitude and phase modulation, there is a competition between the nonlinearity and dispersive factors. After the modulational instability occurs, the growth predicted by linear theory is exponential, and the nonlinear effect in the form of the Akhmediev soliton takes over before the wave trains return to the stage similar to the initial profiles with a phase-shift difference~\cite{kharif2001focusing,slunyaev2002nonlinear}. On the other hand, Biondini and Fagerstrom argued that the major cause of modulational instability in the NLS equation is not the breather soliton solutions per se, but the existence of perturbations where discrete spectra are absence~\cite{biondini2015integrable}.

Experimental attempts on deterministic rogue wave generation using the Akhmediev solitons suggested that the symmetric structure is not preserved and the wave spectrum experiences frequency downshift even though wavefront dislocation and phase singularity are visible~\cite{huijsmans2011experiments,van2011deterministic,karjanto2006mathematical,karjanto2007extreme,karjanto2010qualitative,karjanto2007note}. A numerical calculation of rogue wave composition can be described in the form of the collision of Akhmediev breathers~\cite{akhmediev2009extreme}. Another comparison of the Akhmediev breathers with the North Sea Draupner New Year and the Sea of Japan Yura wave signals also show some qualitative agreement~\cite{chabchoub2010experimental}. The characteristics of the Akhmediev solitons have also been observed experimentally in nonlinear optics~\cite{dudley2009modulation}.

A theoretical, numerical, and experimental report of higher-order modulational instability indicates that a relatively low-frequency modulation on a plane-wave induces pulse splitting at different phases of evolution~\cite{erkintalo2011higher}. Second-order breathers composed of nonlinear combinations of the Kuznetsov-Ma breather and Akhmediev soliton reveal the dependence of the wave envelope on the degenerate eigenvalues and differential shifts~\cite{kedziora2012second}. Similar higher-order Akhmediev solitons visualized in~\cite{erkintalo2011higher,kedziora2012second} has been featured earlier in~\cite{karjanto2006mathematical,karjanto2009mathematical} and similar illustrations can also be found in~\cite{brangerevolution,chabchoub2012super,calini2013observable,onorato2013brogue,ling2013simple,dudley2014instabilities,chabchoub2014time,bilman2019robust}.\\	

\subsection{The Peregrine soliton}
The third one is called the Peregrine soliton, also known as the rational solution~\cite{peregrine1983water}. This soliton is localized in both spatial and temporal variables $(x,t)$ and is written as follows (denoted as $q_\text{P}$):
\begin{equation}
q(x,t) = q_\text{P}(x,t) = e^{2it} \left(\frac{4(1 + 4i t)}{1 + 16 t^2 + 4 x^2} - 1 \right).
\end{equation}
This solution is neither a traveling wave nor contains free parameters. Johnson called it a ``rational-cum-oscillatory solution''~\cite{johnson1997modern}, others referred to it as the ``isolated Ma soliton''~\cite{henderson1999unsteady}, an ``explode-decay solitary wave''~\cite{nakamura1985new}, the ``rational growing-and-decaying mode''~\cite{tajiri1998breather}, the ``algebraic breather''~\cite{yan2010nonautonomous}, or the ``fundamental rogue wave solution''~\cite{ling2013simple}.

The amplitude amplification for the Peregrine soliton is exactly of a factor three and this can be obtained by taking the limit of the parameters toward zero in the previous two breathers:
\begin{equation}
\text{AF}_\text{P} = \lim_{\mu \rightarrow 0} \text{AF}_\text{M} (\mu) = 3 = \lim_{\nu \rightarrow 0} \text{AF}_{A}(\nu).
\end{equation}
Although the other two breather solitons are also proposed as rogue wave prototypes, some authors argued that the Peregrine soliton is the most likely freak wave event due to its appearance from nowhere and disappearance without a trace~\cite{akhmediev2009waves} as well as its closeness to all initial supercritical humps of small uniform envelope amplitude~\cite{shrira2010makes}. Some numerical and experimental studies may support this reasoning.

Henderson et al. studied numerically unsteady surface gravity wave modulations by comparing the fully nonlinear and NLS equations~\cite{henderson1999unsteady}. For steep-wave events, their computations produced striking similarities with the Peregrine soliton. On the other hand, Voronovich et al. confirmed numerically that the bottom friction effect, even when it is small in comparison to the nonlinear term, could hamper the formation of breather freak wave at the nonlinear stage of instability~\cite{voronovich2008can}. Investigations on linear stability demonstrated that the Peregrine soliton is unstable against all standard perturbations, where the analytical study is supported by numerical evidence.~\cite{klein2015numerical,munoz2017instability,calini2019linear,klein2020numerical}.

A sequence of experimental studies using the Peregrine soliton demonstrated reasonably good qualitative agreement with the theoretical prediction. Some discrepancies occur in the modulational gradients, spatiotemporal symmetries, and for larger steepness values~\cite{chabchoub2012experimental}, as well as the frequency downshift~\cite{shemer2013peregrine}. Interestingly, Chabchoub et al. shown further experimentally that the dynamics of the Peregrine soliton and its spectrum characteristics persist even in the presence of wind forcing with high velocity~\cite{chabchoub2013experiments}. By selecting a target location and determining an initial steepness, an experiment using the Peregrine soliton of wave interaction with floating bodies during extreme ocean condition has also been successfully implemented~\cite{onorato2013arogue}.

The Peregrine soliton also finds applications in the evolution of the intrathermocline eddies, also known as the oceanic lenses~\cite{yurova2014hidden}. It appeared as a special case of stationary limit in the solutions of the spinor BEC model~\cite{li2018solitons}, and it was observed experimentally emerging from the stochastic background in surface gravity deep-water waters~\cite{cazaubiel2018coexistence}.

Nonlinear spectral analysis using the finite gap theory showed that the spectral portraits of the Peregrine soliton represent a degenerate genus two of the NLS equation solution~\cite{randoux2018nonlinear}. Higher-order Peregrine solitons in terms of quasi-rational functions are derived in~\cite{gaillard2015multi}. Higher-order Peregrine solitons up to the fourth-order using a modified Darboux transformation has been presented with applications in rogue waves in the deep ocean and high-intensity rogue light wave pulses in optical fibers~\cite{akhmediev2009rogue}. Super rogue waves modeled with higher-order Peregrine soliton with an amplitude amplification factor of five times the background value are observed experimentally in a water-wave tank~\cite{chabchoub2012super}. 

The following section will show that as the parameter values $\mu \rightarrow 0$ and $\nu \rightarrow 0$, the Kuznetsov-Ma and Akhmediev breathers reduce to the Peregrine soliton.

\section{Limiting behavior} 		\label{limitbehave}

This section provides a rigorous proof of the limiting behavior of breather wave solutions using the $\epsilon$-$\delta$ argument. We have the following theorem:
\begin{theorem}
The Peregrine soliton is a limiting case for both the Kuznetsov-Ma breather and Akhmediev soliton:
\begin{equation}
\lim_{\mu \rightarrow 0} q_{\textmd{M}}(x,t) = q_{\textmd{P}}(x,t) = \lim_{\nu \rightarrow 0} q_{\textmd{A}}(x,t).
\end{equation}
\end{theorem}
We divide the proof into four parts, and each limit consists of two parts corresponding to the real and imaginary parts of the solitons.
\begin{proof}
The following shows that the limit for the real parts of the Kuznetsov-Ma breather and Peregrine soliton is correct, i.e.,
\begin{equation*}
\lim_{\nu \rightarrow 0} \text{Re} \left\{ q_{\textmd{M}}(x,t) \right\} = \text{Re} \left\{ q_{\text{P}}(x,t) \right\}.
\end{equation*}
For each $\epsilon > 0$, there exists $\delta = \sqrt{(\epsilon + 2)^2 - 4} > 0$ such that if $0 < \mu < \delta$, then $|\text{Re} \left\{q_\text{M}\right\} - \text{Re} \left\{q_\text{P} \right\} | < \epsilon$.
We know that since 
\begin{equation*}
\frac{\cosh \mu (x - x_0)}{\cos \rho (t - t_0)} \geq 1 \qquad \text{for all} \; (x,t) \in \mathbb{R}^2
\end{equation*}
it then implies
\begin{equation*}
\rho \frac{\cosh \mu (x - x_0)}{\cos \rho (t - t_0)} - 2 \mu \geq \rho - 2 \mu.
\end{equation*}
It follows that
\begin{align*}
|\text{Re} \left\{q_\text{M}\right\} - \text{Re} \left\{q_\text{P} \right\} | 
&= \left|\frac{\mu^3}{\rho \frac{\cosh \nu (x - x_0)}{\cos \rho (t - t_0)} - 2\mu} - \frac{4}{1 + 16(t - t_0)^2 + 4(x - x_0)^2} \right| \\
&\leq \left|\frac{\mu^3}{\rho - 2\mu} - 4 \right| = \left|\sqrt{\mu^2 + 4} - 2 \right| \\
&\leq \sqrt{\delta^2 + 4} - 2 = \sqrt{ \left(\sqrt{(\epsilon + 2)^2 - 4} \right)^2 + 4} - 2 = \epsilon.
\end{align*}

The following verifies that the limit for the imaginary parts of the Kuznetsov-Ma breather and Peregrine soliton is accurate, i.e.,
\begin{equation*}
\lim_{\nu \rightarrow 0} \text{Im} \left\{ q_{\textmd{M}}(x,t) \right\} = \text{Im} \left\{ q_{\text{P}}(x,t) \right\}.
\end{equation*}
For each $\epsilon > 0$, there exists $\delta = \sqrt{-10 + 2 \sqrt{25 + 4\epsilon/|t - t_0|}} > 0$ such that if $0 < \mu < \delta$, then 
$|\text{Im} \left\{q_\text{M} \right\} - \text{Im} \left\{q_\text{P} \right\} | < \epsilon$.
We can write the imaginary parts of $q_\text{M}$ and $q_\text{P}$ as follows
\begin{align*}
\text{Im}\left\{q_\text{M} \right\} &= \frac{\mu \rho}{\rho \frac{\cosh \mu(x - x_0)}{\sin \rho (t - t_0)} - 2 \mu \cot \rho(t - t_0)}
\leq \frac{\mu \rho^2 |t - t_0|}{\rho - 2 \mu} \\
\text{Im}\left\{q_\text{P} \right\} &= \frac{16(t - t_0)}{1 + 16 (t - t_0)^2 + 4(x - x_0)^2} \leq 16 |t - t_0|.
\end{align*}
It follows that
\begin{align*}
\left|\text{Im}\left\{q_\text{M} \right\} - \text{Im}\left\{q_\text{P} \right\} \right|
&= \left|\frac{\mu \rho}{\rho \frac{\cosh \mu(x - x_0)}{\sin \rho (t - t_0)} - 2 \mu \cot \rho(t - t_0)} - \frac{16(t - t_0)}{1 + 16 (t - t_0)^2 + 4(x - x_0)^2} \right| \\
&\leq \left|\frac{\mu \rho^2}{\rho - 2 \mu} - 16 \right|  |t - t_0| 
= \left|(\mu^2 + 4) \left(\sqrt{\mu^2 + 4} + 2 \right) - 16 \right| |t - t_0| \\
&< \left|\left(\delta^2 + 4 \right) \left(\frac{\delta^2}{4} + 4 \right) - 16 \right| |t - t_0| 
= \left|\frac{\delta^4}{4} + 5 \delta^2 \right| |t - t_0| = \epsilon.
\end{align*}
	
In what follows, we present the limit of the real part of the Akhmediev soliton as $\nu \rightarrow 0$ is indeed the real part of the Peregrine solitons, i.e.,
\begin{equation*}
\lim_{\nu \rightarrow 0} \text{Re} \left\{ q_{\textmd{A}}(x,t) \right\} = \text{Re} \left\{ q_{\text{P}}(x,t) \right\}.
\end{equation*}
For each $\epsilon > 0$, there exists $\delta = \sqrt{\epsilon/3} > 0$ such that if $0 < \nu < \delta < 2$, then 
$|\text{Re} \left\{q_\text{A} \right\} - \text{Re} \left\{q_\text{P} \right\} | < \epsilon$.
We know that since 
\begin{equation*}
-1 \leq \frac{\cos \nu (x - x_0)}{\cosh \sigma (t - t_0)} \leq 1 \qquad \text{for all} \; (x,t) \in \mathbb{R}^2
\end{equation*}
it then implies
\begin{equation*}
2\nu - \sigma \leq 2 \nu - \sigma \frac{\cos \nu (x - x_0)}{\cosh \sigma (t - t_0)}.
\end{equation*}
We also have $1 + 16 (t - t_0)^2 + 4(x - x_0)^2 \geq 1$ for all $(x,t) \in \mathbb{R}^2$.
Furthermore, since $0 \leq \sqrt{4 - \nu^2} \leq 2$, $0 \leq 2 - \sqrt{4 - \nu^2} \leq 2$,
\begin{align*}
                  \frac{1}{4} \leq \frac{2 - \sqrt{4 - \nu^2}}{\nu^2} &\leq \frac{1}{2},	 \qquad 
\qquad  \frac{1}{2} \leq \frac{1}{4} + \frac{2 - \sqrt{4 - \nu^2}}{\nu^2} \leq \frac{3}{4} \\
\text{and} \qquad \frac{2 - \sqrt{4 - \nu^2}}{4\nu^2} &\geq \frac{1}{4\delta^2}
\end{align*}
it follows that
\begin{align*}
|\text{Re} \left\{q_\text{A}\right\} - \text{Re} \left\{q_\text{P} \right\} | 
&= \left|\frac{\nu^3}{2 \nu - \sigma \frac{\cos \nu (x - x_0)}{\cosh \sigma (t - t_0)}} - \frac{4}{1 + 16(t - t_0)^2 + 4(x - x_0)^2} \right| \\
&\leq \left|\frac{\nu^3}{2 \nu - \sigma} + 4 \right| = \left|\frac{1}{\frac{2 - \sqrt{4 - \nu^2}}{\nu^2}} + \frac{1}{\frac{1}{4}} \right| 
= \frac{\left|\frac{1}{4} + \frac{2 - \sqrt{4 - \nu^2}}{\nu^2} \right|}{\left|\frac{2 - \sqrt{4 - \nu^2}}{4\nu^2} \right|} \\
&\leq \frac{3}{4} \left(4 \delta^2 \right) = 3 \left(\sqrt{\frac{\epsilon}{3}} \right)^2 = \epsilon.
\end{align*}

In what follows, we demonstrate that the limit of the imaginary part of the Akhmediev soliton becomes the imaginary part of the Peregrine soliton, i.e.,
\begin{equation*}
\lim_{\nu \rightarrow 0} \text{Im} \left\{ q_{\textmd{A}}(x,t) \right\} = \text{Im} \left\{ q_{\text{P}}(x,t) \right\}.
\end{equation*}
For each $\epsilon > 0$, there exists $\delta = \sqrt{\epsilon/4} > 0$ such that if $0 < \nu < \delta < 2$, then 
$|\text{Im} \left\{q_\text{A} \right\} - \text{Im} \left\{q_\text{P} \right\} | < \epsilon$.
We can write the imaginary parts of $q_\text{A}$ and $q_\text{P}$ as follows
\begin{align*}
\text{Im}\left\{q_\text{A} \right\} &= \frac{\nu \sigma \tanh \sigma (t - t_0)}{2 \nu - \sigma \frac{\cos \nu(x - x_0)}{\cosh \sigma (t - t_0)}}
\leq \frac{\nu \sigma^2 |t - t_0|}{2\nu - \sigma} \\
\text{Im}\left\{q_\text{P} \right\} &= \frac{16(t - t_0)}{1 + 16 (t - t_0)^2 + 4(x - x_0)^2} \leq 16 |t - t_0|.
\end{align*}
Since $2 + \sqrt{4 - \nu^2} \leq 4$ and $(4 - \nu^2) \leq 4 + \delta^2/|t - t_0|$, it follows that
\begin{align*}
\left|\text{Im}\left\{q_\text{A} \right\} - \text{Im}\left\{q_\text{P} \right\} \right| 
&= \left|\frac{\nu \sigma \tanh \sigma (t - t_0)}{2 \nu - \sigma \frac{\cos \nu(x - x_0)}{\cosh \sigma (t - t_0)}} - \frac{16(t - t_0)}{1 + 16 (t - t_0)^2 + 4(x - x_0)^2} \right| 
\end{align*} 
\begin{align*}
&\leq \left|\frac{\nu \sigma^2}{2\nu - \sigma} - 16 \right| \left|t - t_0\right| 
 = \left|(4 - \nu^2)(2 + \sqrt{4 - \nu^2}) - 16 \right| |t - t_0| \\
&< \left| 4 \left(4 + \frac{\delta^2}{|t - t_0|} \right) - 16 \right| |t - t_0| = 4 \delta^2 = 4 \left(\sqrt{\frac{\epsilon}{4}}\right)^2 = \epsilon.
\end{align*} 
We have completed the proof. \hfill $\square$
\end{proof}
In the following section, we will visualize the limiting behavior of the breather solutions as they approach toward the Peregrine soliton.

\section{Limiting behavior visualized}			\label{libevisual}

\begin{figure}[h!]
\begin{center}
\includegraphics[width=0.9\textwidth]{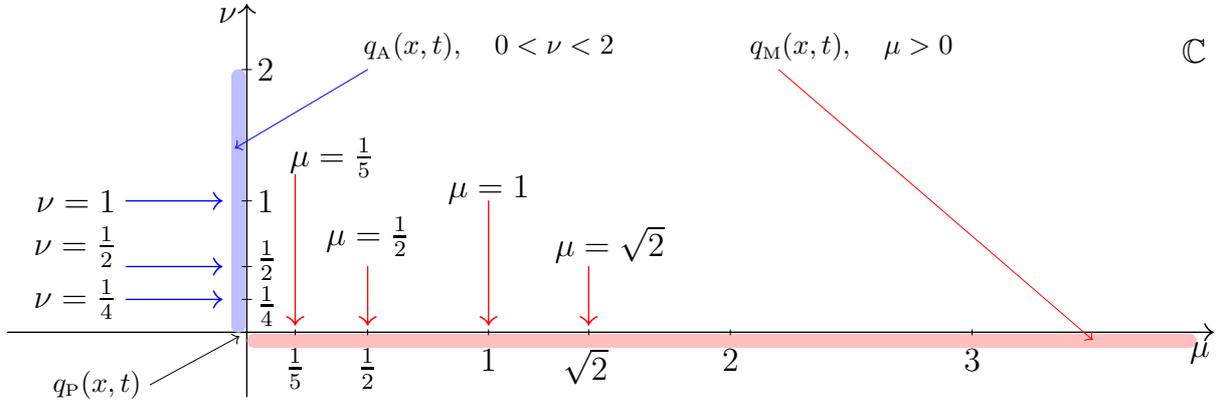}
\end{center}
\caption{Selected parametric values $\nu$ and $\mu = i\nu$ displayed in the complex plane for the Kuznetsov-Ma breather and Akhmediev soliton visualized in this section.} \label{fig1-munu}
\end{figure}

In this section, we will visually confirm the limiting behavior of the Kuznetsov-Ma and Akhmediev breathers toward the Peregrine soliton as both parameter values approach zero. Subsection~\ref{contourplot} presents the contour plots of the amplitude modulus, and Subsection~\ref{parameterization} discusses the spatial and temporal parameterizations of the breathers. We select several parameter values in sketching the plots. Figure~\ref{fig1-munu} displays the chosen parametric values for both breather solutions, where they can be visualized in the complex-plane for the parameter pair $(\mu,\nu)$.\\

\subsection{Contour plot}	\label{contourplot}

In this subsection, we observe the contour plots of the amplitude modulus of the breather and how the changes in the parameter values affect the envelope's period and wavelength. Similar contour plots have been presented in the context of electronegative plasmas with Maxwellian negative ions~\cite{el2017nonlinear}. In particular, the contour plot of the Peregrine soliton is also displayed in~\cite{chabchoub2012experimental}.

\begin{figure}[h!]
\begin{center}
\subcaptionbox{(a) $\mu = \sqrt{2}$}{\includegraphics[width=0.3\textwidth]{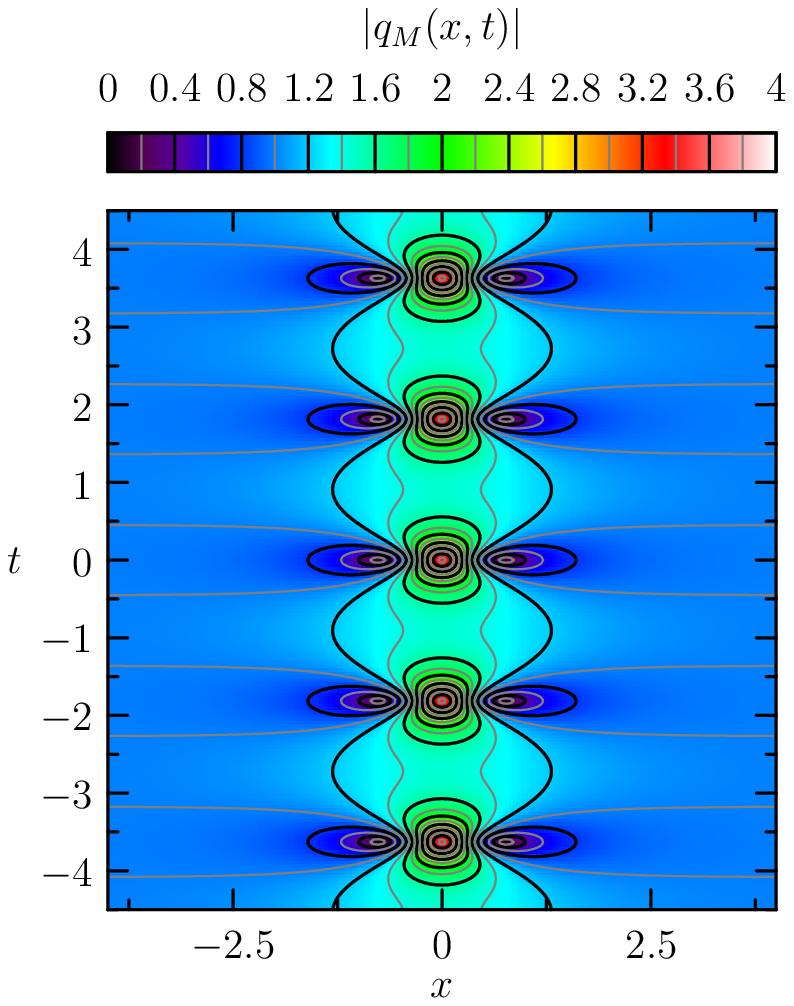}}			\hspace*{0.5cm}
\subcaptionbox{(b) $\mu = 1$}   {\includegraphics[width=0.3\textwidth]{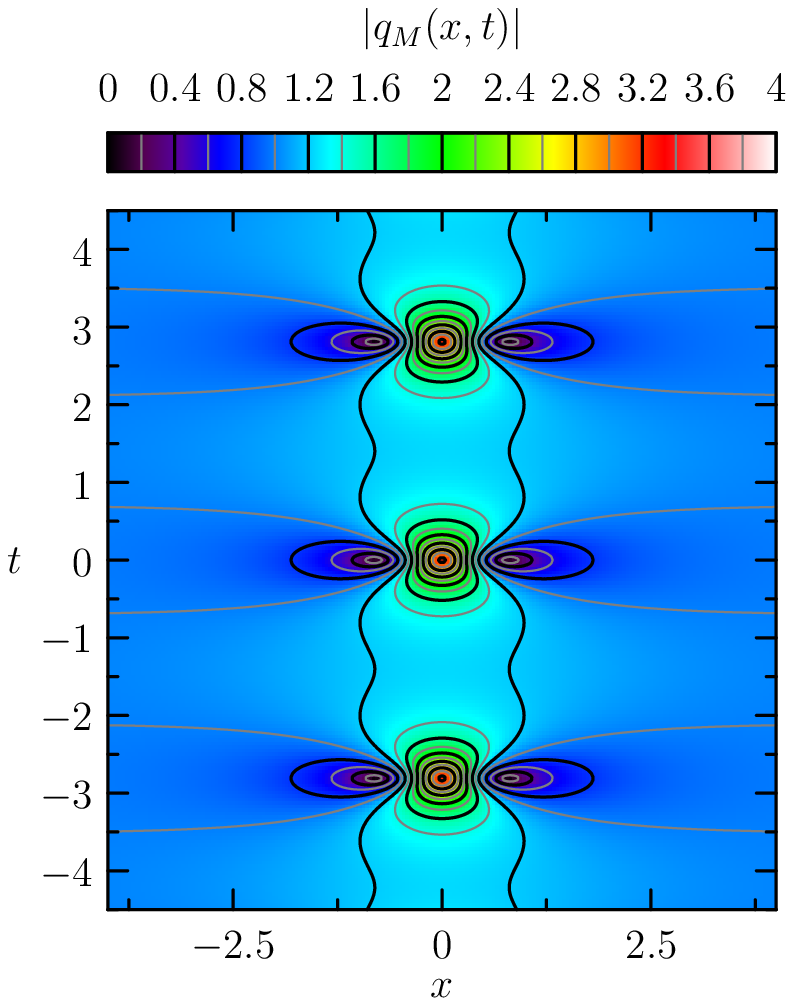}}				\hspace*{0.5cm}
\subcaptionbox{(c) $\mu = \frac{1}{2}$} {\includegraphics[width=0.3\textwidth]{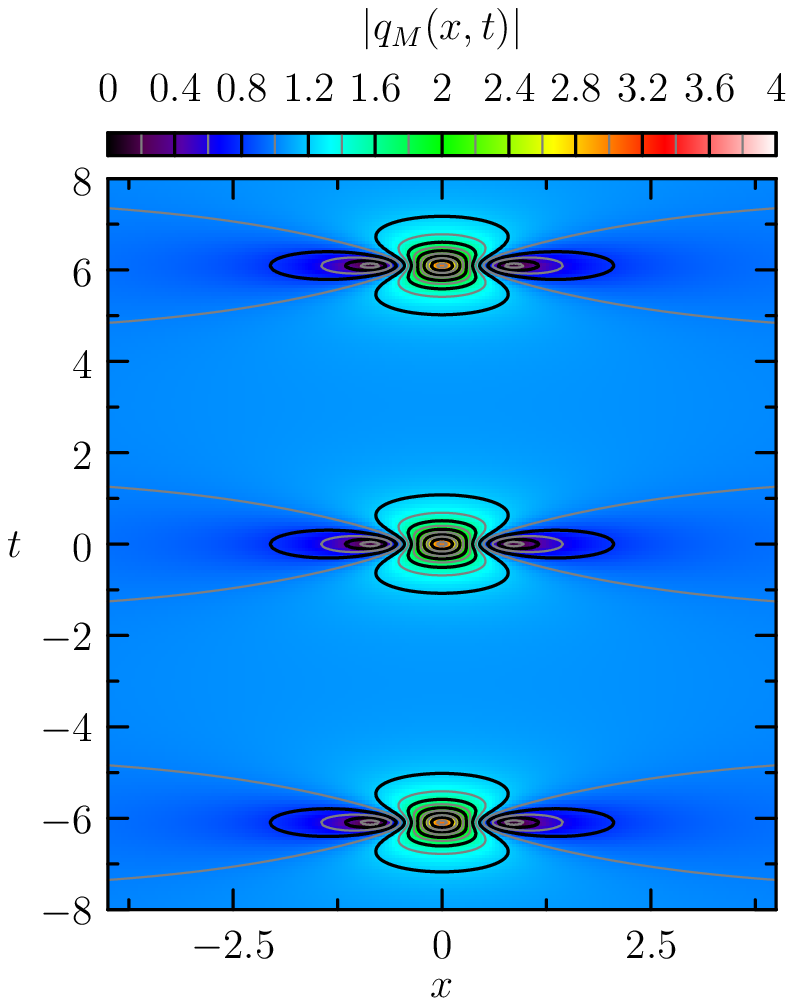}}			
\subcaptionbox{(d) $\mu = \frac{1}{5}$} {\includegraphics[width=0.3\textwidth]{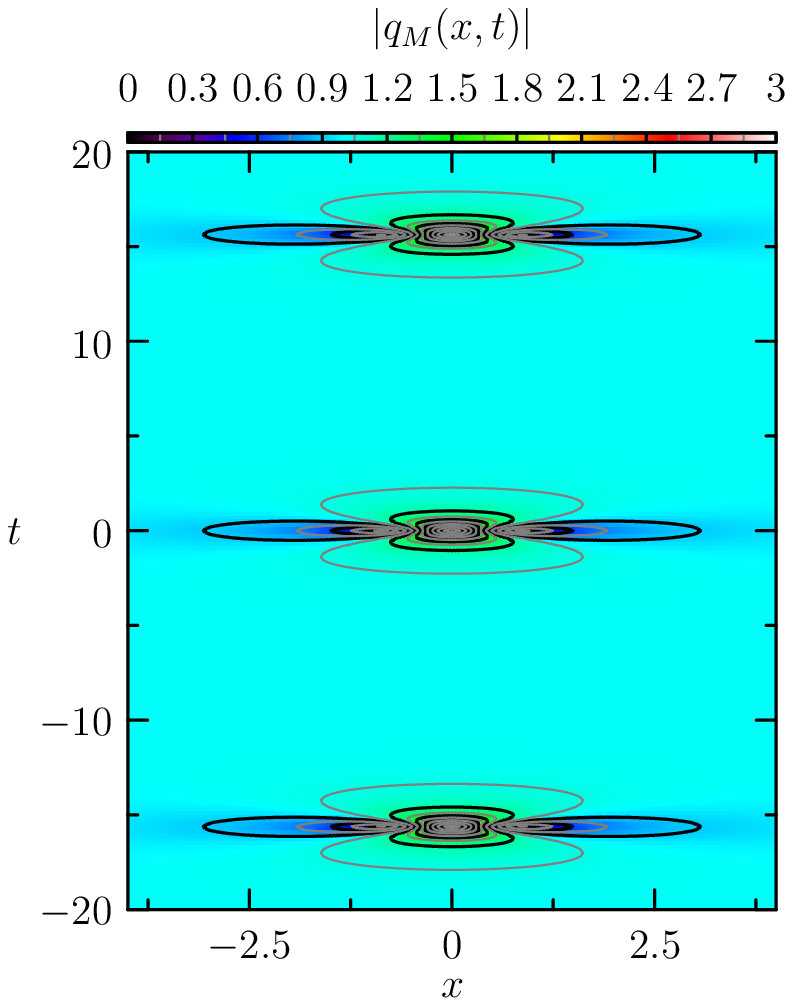}}				\hspace*{0.5cm}
\subcaptionbox{(e) $\mu = \frac{1}{5}$} {\includegraphics[width=0.3\textwidth]{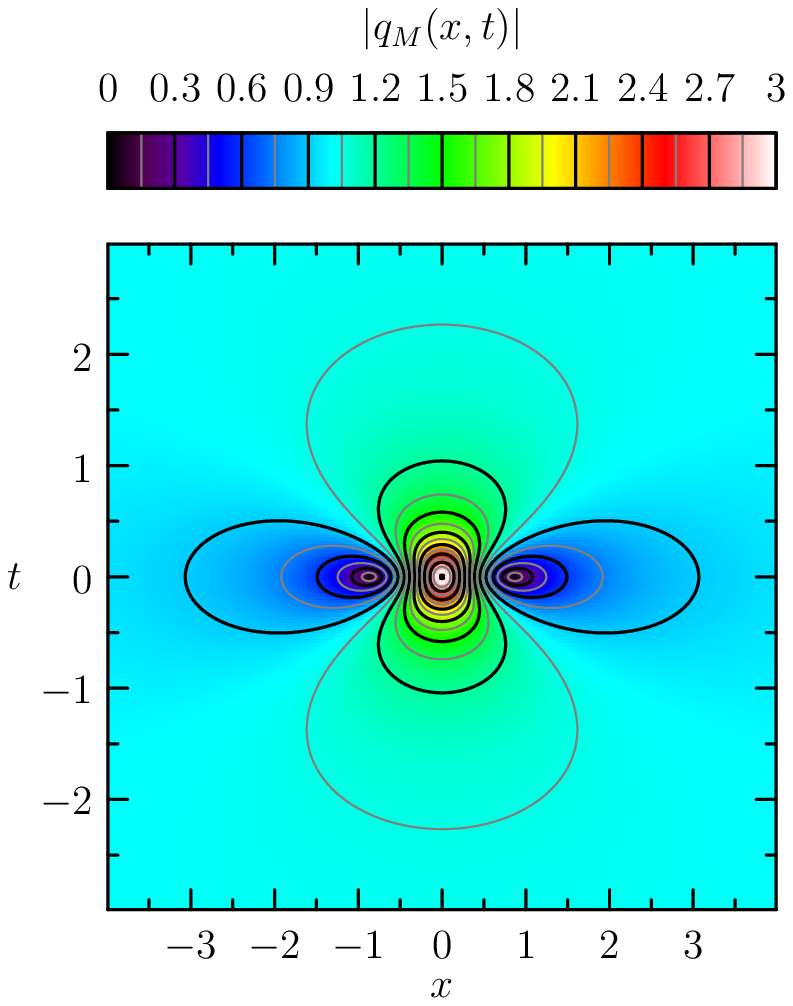}}				\hspace*{0.5cm}
\subcaptionbox{(f) $\mu = 0$}{\includegraphics[width=0.3\textwidth]{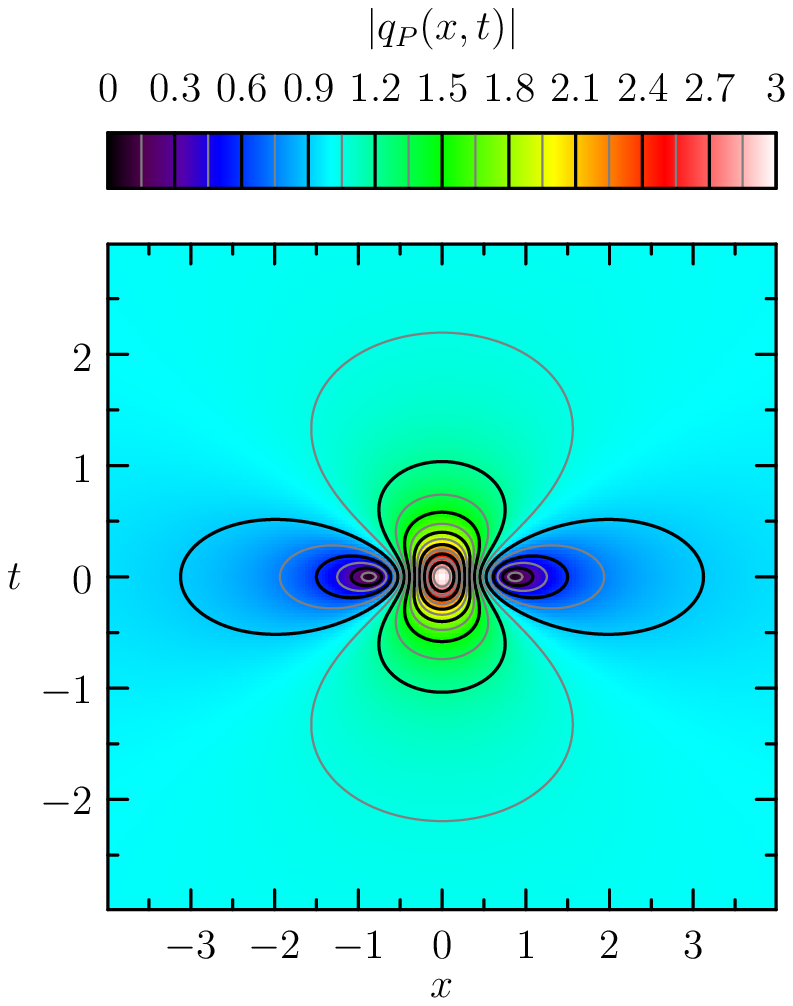}}	
\end{center}
\caption{Contour plots for the moduli of the Kuznetsov-Ma breather for (a) $\mu = \sqrt{2}$, (b) $\mu = 1$, (c) $\mu = 0.5$, (d) $\mu = 0.2$, (e) also $\mu = 0.2$ but a zoom-in version, and (f) $\mu = 0$, which gives the Peregrine soliton. Notice that the contour plots (e) and (f) are qualitatively nearly identical.} \label{fig2-kuzma-perat-contour}
\end{figure}
Figures~\ref{fig2-kuzma-perat-contour}(a)--\ref{fig2-kuzma-perat-contour}(e) display the contour plots of the Kuznetsov-Ma breather for several values of parameters $\mu$: $\sqrt{2}, 1, \frac{1}{2}$, and $\frac{1}{5}$. Figure~\ref{fig2-kuzma-perat-contour}(e) is a zoom-in version of the same contour plot given in Figure~\ref{fig2-kuzma-perat-contour}(d). Figure~\ref{fig2-kuzma-perat-contour}(f) is the final stop when we let the parameter $\mu \rightarrow 0$, for which the Kuznetsov-Ma breather turns into the Peregrine soliton. It is interesting to note that for $\mu = \frac{1}{5}$, the contour plot is nearly identical with the one from the Peregrine soliton, as we can observe by qualitatively comparing panels~(e) and~(f) of Figure~\ref{fig2-kuzma-perat-contour}. 

\begin{table}[h]
\begin{center}
\begin{tabular}{@{}ccccccc@{}}	
\toprule
\multicolumn{4}{c}{Parameter values} &  &\multicolumn{2}{c}{Temporal envelope period} \\
\cline{1-4} \cline{6-7} 
$\mu$ (exact)    & $\mu$ (decimal)  & $\rho$ (exact)	& $\rho$ (approximation) && $T_\text{M}$ (exact) 	  & $T_\text{M}$ (approximation)   \\ \hline 
$1/5$            & 0.2       		& $\sqrt{101}/25$	& $0.402$				 && $50\pi/\sqrt{101}$ & $15.630$  				\\ 
$1/2$            & 0.5       		& $\sqrt{17}/4$		& $1.031$				 && $8\pi/\sqrt{17}$   &  $6.096$  				\\ 
$1$ 	         & 1.0		 		& $ \sqrt{5}$		& $2.236$				 && $2\pi/\sqrt{5}$    &  $2.810$ 				\\ 
$\sqrt{2}$	 	 & 1.414	 		& $2\sqrt{3}$		& $3.464$				 && $\pi/\sqrt{3}$     &  $1.814$   				\\ 
\bottomrule
\end{tabular}
\end{center}
\caption{Exact values of the temporal envelope period $T_\text{M}$ and their approximate values for selected parameter values $\mu$ corresponding to the Kuznetsov-Ma breather.} \label{table1-kuzma}
\end{table}
Let $T_\text{M}$ denote the temporal envelope period for the Kuznetsov-Ma breather, then we know that in general, $T_\text{M} = 2\pi/\rho$. For $\mu \rightarrow \infty$, $T_\text{M} \rightarrow 0$ and vice versa, for $\mu \rightarrow 0$, $T_\text{M} \rightarrow \infty$. For any given value of $\mu > 0$, $T_\text{M}$ can be easily calculated. Here are some examples. For $\mu = \sqrt{2}$, $T_\text{M} = \pi/\sqrt{3} \approx 1.814$ and we display five periods in Figure~\ref{fig2-kuzma-perat-contour}(a) along the temporal axis~$t$. For $\mu = 1$, $T_\text{M} = 2\pi/\sqrt{5} \approx 2.81$ and for the same time interval as in panel~(a), we can only capture three periods along the temporal axis $t$, as shown in Figure~\ref{fig2-kuzma-perat-contour}(b). Furthermore, for $\mu = \frac{1}{2}$, $T_\text{M} = 8\pi/\sqrt{17} \approx 6.1$ and we need to extend almost twice length in the time interval in order to capture at least three periods. Figure~\ref{fig2-kuzma-perat-contour}(c) shows this contour plot. Finally, for $\mu = \frac{1}{5}$, $T_\text{M} = 50\pi/\sqrt{101} \approx 15.63$. As we can observe in Figure~\ref{fig2-kuzma-perat-contour}(d), extending the length of time interval to around 40 units is sufficient to capture at least three periods, albeit the detail around maximum and minimum is hardly visible. Table~\ref{table1-kuzma} displays selected parameter values of the Kuznetsov-Ma breather and their corresponding temporal envelope periods~$T_\text{M}$. 

\begin{figure}[h!]
\begin{center}
\subcaptionbox{(a) $\nu = 1$}	{\includegraphics[width=0.45\textwidth]{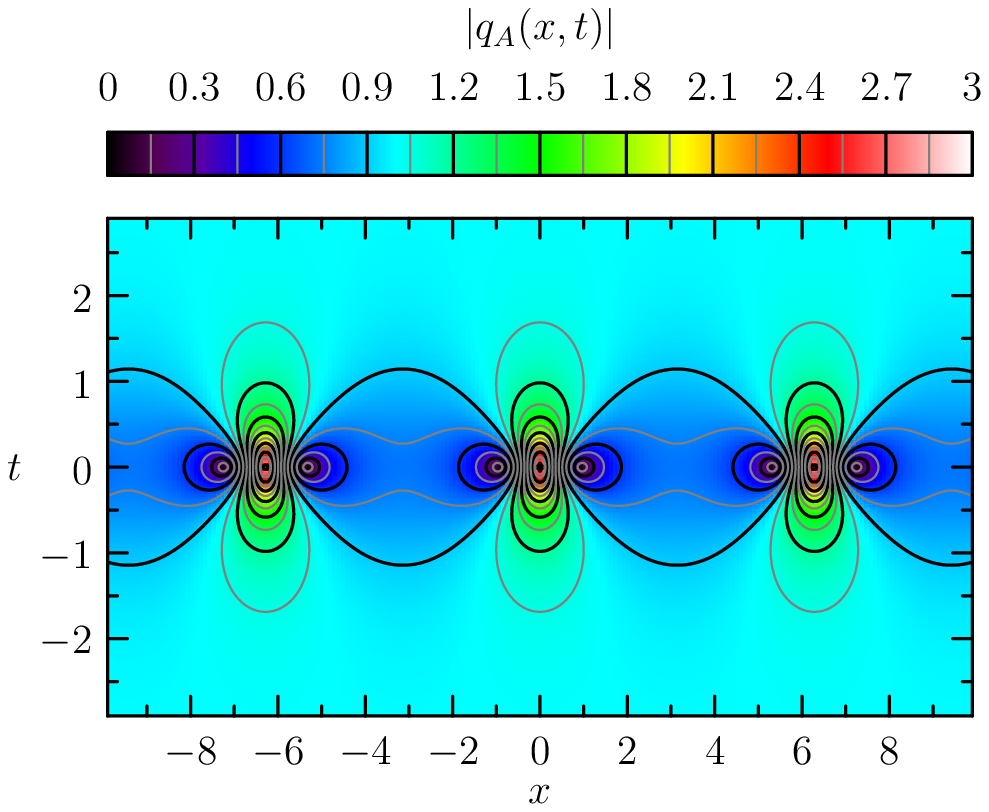}}		\hspace*{1cm}
\subcaptionbox{(b) $\nu = \frac{1}{2}$} {\includegraphics[width=0.45\textwidth]{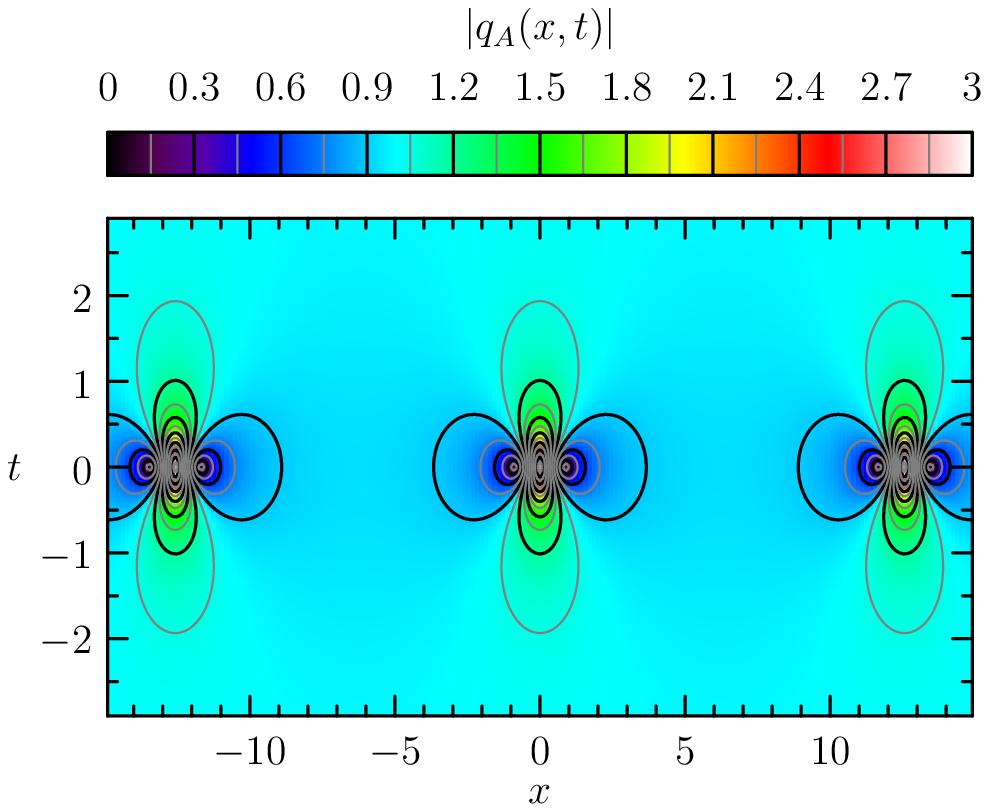}}
\subcaptionbox{(c) $\nu = \frac{1}{4}$}{\includegraphics[width=0.45\textwidth]{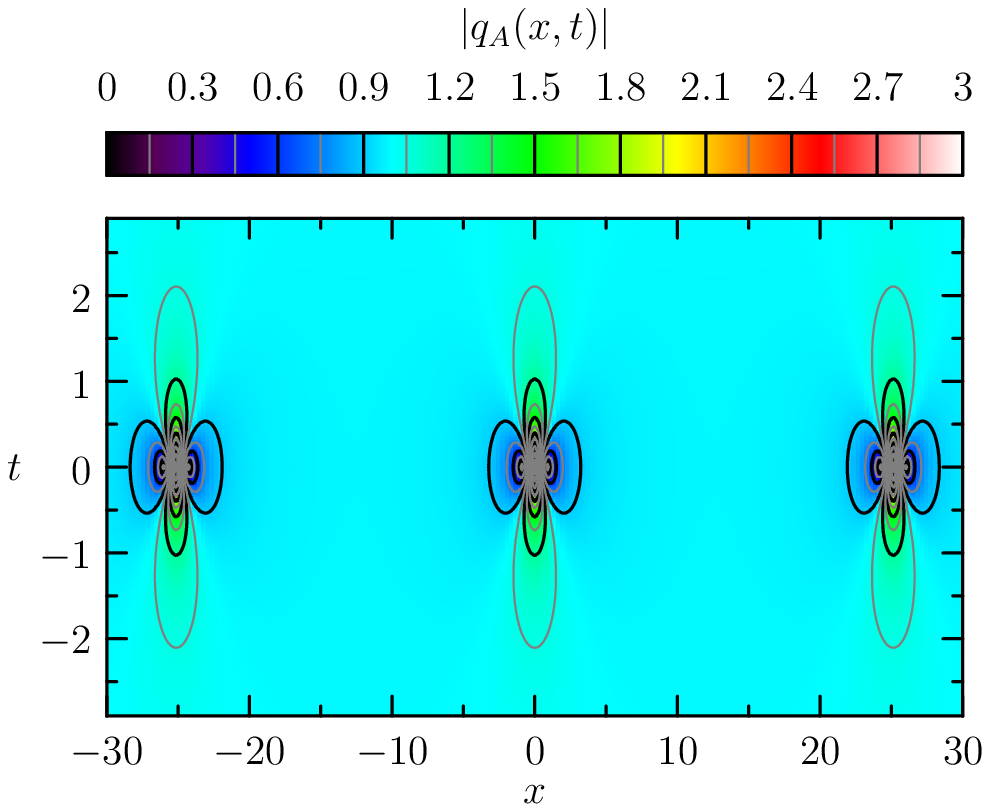}}		\hspace*{1cm}
\subcaptionbox{(d) $\nu = 0$}{\includegraphics[width=0.425\textwidth]{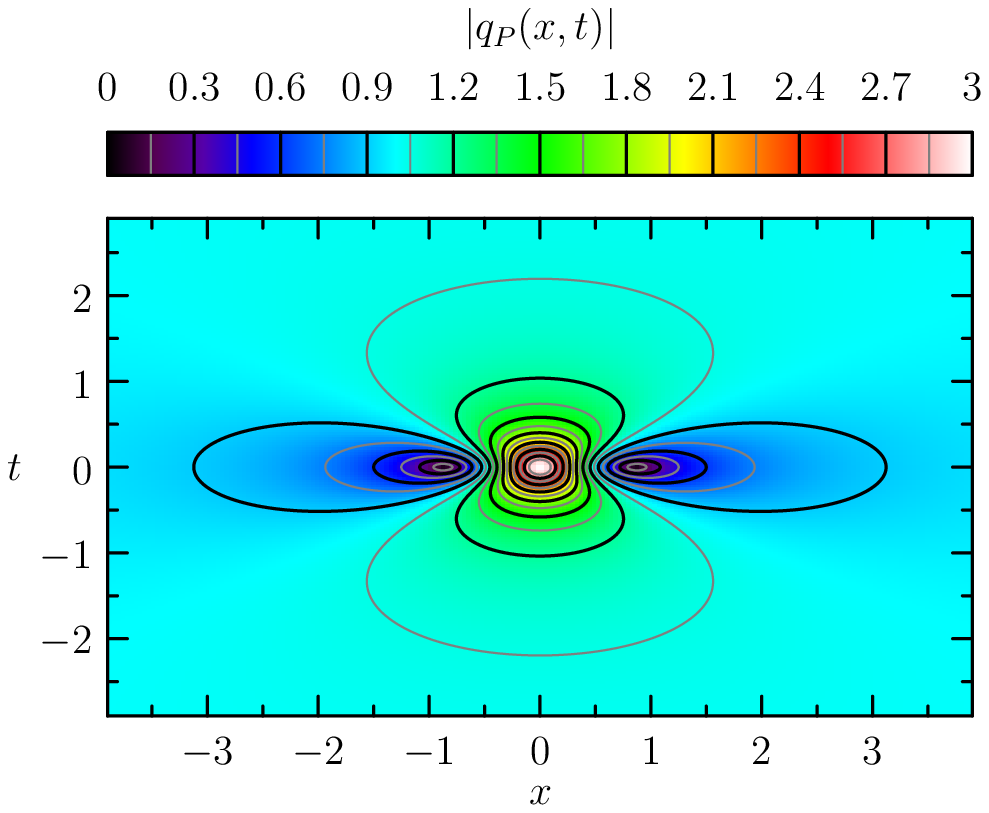}}
\end{center}
\caption{Contour plots for the moduli of the Akhmediev breather for (a) $\nu = 1$, (b) $\nu = 0.5$, and (c) $\nu = 0.25$, as well as (d) the Peregrine soliton.} \label{fig3-akmi-perat-contour}	\vspace*{-0.25cm}
\end{figure}
Figures~\ref{fig3-akmi-perat-contour}(a)--\ref{fig3-akmi-perat-contour}(c) display the contour plot of the Akhmediev soliton for selected values of its parameters $\nu$: $1$, $\frac{1}{2}$, and $\frac{1}{4}$. Figure~\ref{fig3-akmi-perat-contour}(d) shows the contour plot of the Peregrine soliton, which occurs as the final destination when letting the parameter $\nu \rightarrow 0$. Figure~\ref{fig3-akmi-perat-contour}(d) is identical to Figure~\ref{fig2-kuzma-perat-contour}(f), the only difference lies in the length-scale of both horizontal and vertical axes. Similar to the previous case, zooming-in the contour plot for $\nu = \frac{1}{4}$ in Figure~\ref{fig3-akmi-perat-contour}(c) will yield a qualitatively nearly identical contour plot with the Peregrine soliton shown in the panel~(d). (It is not shown in the figure.) \vspace*{-0.25cm}

\begin{table}[h!]
\begin{center}
\begin{tabular}{@{}ccccccc@{}}	
\toprule
\multicolumn{4}{c}{Parameter values} &  &\multicolumn{2}{c}{Spatial envelope wavelength} \\
\cline{1-4} \cline{6-7} 
$\nu$ (exact)    & $\nu$ (decimal)  & $\sigma$ (exact)	& $\sigma$ (approximation) && $L_\text{A}$ (exact) 	  & $L_\text{A}$ (approximation)   \\ \hline 
$1/4$            & 0.25       		& $3\sqrt{7}/16$	& $0.496$				   && $8\pi$ 			  & $25.133$  				\\ 
$1/2$            & 0.5       		& $\sqrt{15}/4$		& $0.968$				   && $4\pi$   			  & $12.566$  				\\ 
$1$ 	         & 1.0		 		& $\sqrt{3}$		& $1.732$				   && $2\pi$		      &  $6.283$ 				\\ 
\bottomrule
\end{tabular}
\end{center}
\caption{Exact values of the spatial envelope wavelength $L_\text{A}$ and their approximate values for selected parameter values $\nu$ corresponding to the Akhmediev soliton.} \label{table2-akmi}
\end{table}
Let $L_\text{A}$ denote the spatial envelope wavelength for the Akhmediev soliton, then for $0 < \nu < 2$, $L_\text{A} = 2\pi/\nu$, which gives $L_\text{A} > \pi$. For $\nu \rightarrow 2$, $L_\text{A} \rightarrow \pi$ and for $\nu \rightarrow 0$, $L_\text{A} \rightarrow \infty$. Table~\ref{table2-akmi} displays selected values of the parameter $\nu$ and their corresponding spatial envelope wavelength $L_\text{A}$ for the Akhmediev soliton. For $\nu = 1$, $L_\text{A} = 2\pi$ and the spatial length of 20 units in Figure~\ref{fig3-akmi-perat-contour}(a) is sufficient to capture three envelope wavelength. For $\nu = 1/2$, $L_\text{A} = 4\pi$ and the spatial length of 40 units in Figure~\ref{fig3-akmi-perat-contour}(b) is required to capture at least three envelope wavelength. For $\nu = 1/4$, $L_\text{A} = 8\pi$ and the spatial length of 60 units in Figure~\ref{fig3-akmi-perat-contour}(c) is needed to capture at least three envelope wavelength. The details around maxima and minima are hardly visible for the latter. 

\subsection{Parameterization in spatial and temporal variables}			\label{parameterization}

In this subsection, we write the breather solutions as $q_\text{X}(x,t) = q_0(t) \, \tilde{q}_\text{X}(x,t)$, where $q_0(t)$ is the plane-wave solution and $\text{X} = \left\{\text{M}, \text{A}, \text{P} \right\}$. Since the plane-wave solution gives a fast-oscillating effect, we only consider the non-rapid oscillating part of the breathers $\tilde{q}_\text{X}$ for the parameterization visualization. In the subsequent figures, we present both spatial and temporal parameterizations of the Kuznetsov-Ma breather, Akhmediev, and Peregrine solitons. 
A similar description has been briefly covered and discussed in~\cite{akhmediev1992phase,akhmediev1997solitons,chabchoub2014hydrodynamics,kimmoun2016modulation}.

\begin{figure}[htbp!]
\begin{center}
\subcaptionbox{(a) $\mu = \sqrt{2}$}{\includegraphics[width=0.4\textwidth]{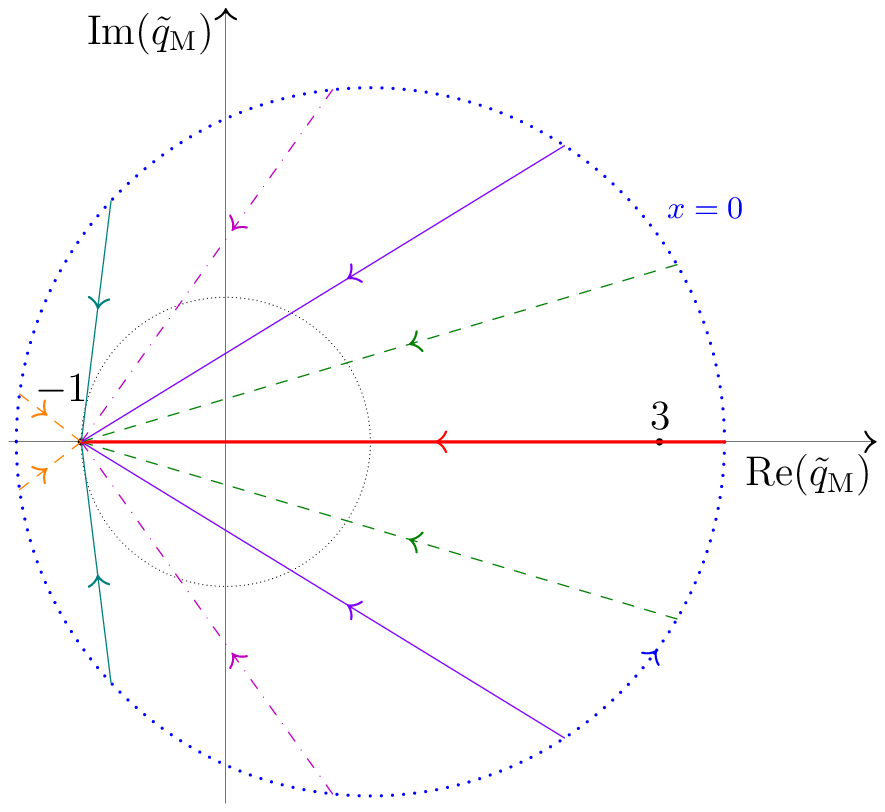}}		\hspace*{1cm}
\subcaptionbox{(b) $\mu = 1$}   	{\includegraphics[width=0.4\textwidth]{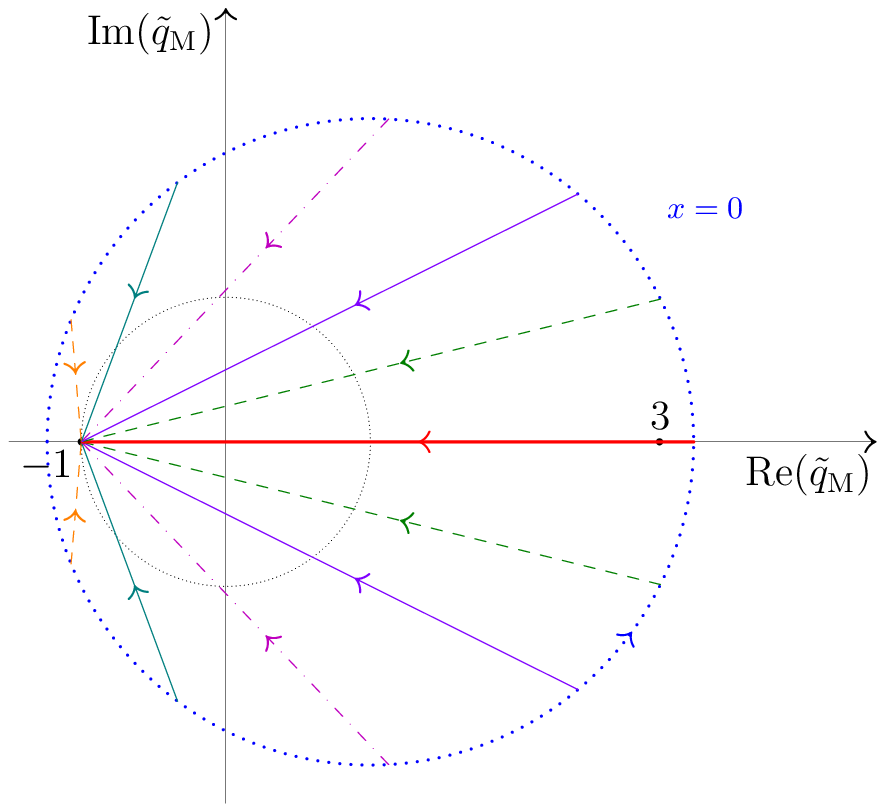}}			
\subcaptionbox{(c) $\mu = 1/2$}		{\includegraphics[width=0.4\textwidth]{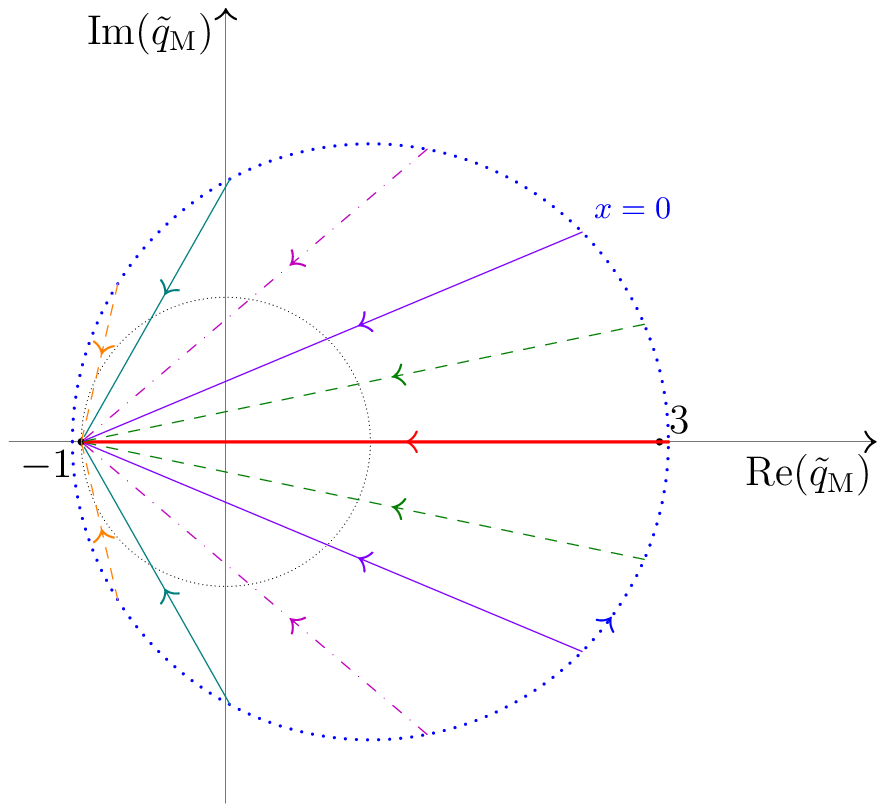}}		\hspace*{1cm}
\subcaptionbox{(d) $\mu = 1/5$}     {\includegraphics[width=0.4\textwidth]{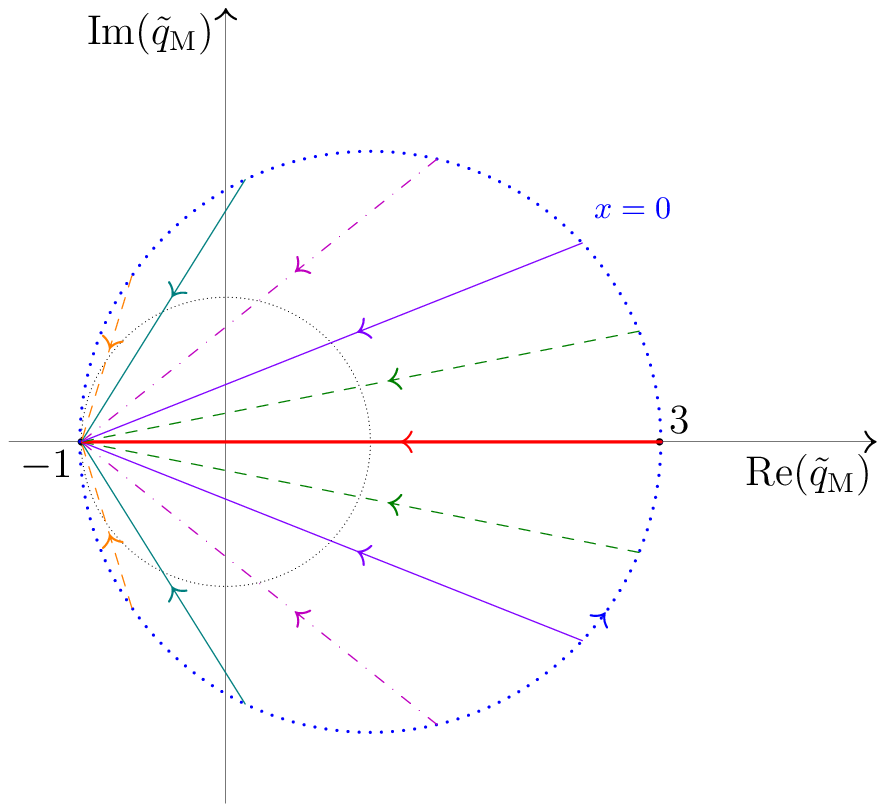}}		
\subcaptionbox{}{\includegraphics[width=0.9\textwidth]{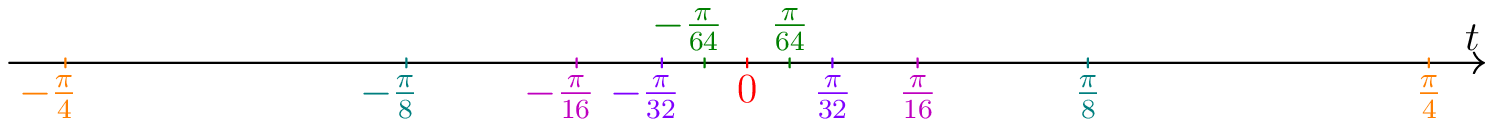}}
\end{center}
\caption{Parameterization of the non-rapid-oscillating complex-valued amplitude of the Kuznetsov-Ma breather $\tilde{q}_\text{M}$ in the spatial variable $x$, $x \geq 0$, for different values of temporal variable $t$ and different values of the parameter $\mu$: (a) $\mu = \sqrt{2}$, (b) $\mu = 1$, (c) $\mu = 1/2$, and (d) $\mu = 1/5$. The selected values of $t$ are $t = 0$ (solid red), $t = \pm \pi/64$ (dashed green), $t = \pm \pi/32$ (solid purple), $t = \pm \pi/16$ (dash-dotted magenta), $t = \pm \pi/8$ (solid cyan), and $t = \pm \pi/4$ (dashed orange).} \label{fig4-argkmx-param}
\end{figure}
Figure~\ref{fig4-argkmx-param} displays the parameterization of the non-rapid oscillating Kuznetsov-Ma breather $\tilde{q}_\text{M}$ in the spatial variable $x$ for different values of the temporal variable $t$ and parameter $\mu$. Different panels indicate different parameter values $\mu$ and for each panel, different curves, for which in this particular case, they are merely straight lines, indicate different time $t$. For all cases, we consider $x \geq 0$ due to the symmetry nature of the breathers. The straight-line trajectories move inwardly focused from the dotted blue circle at $x = 0$ toward $(-1,0)$ as $x \rightarrow \infty$. The situation is simply reversed for $x < 0$: the path of trajectories move outwardly defocused as $x$ progresses from $(-1,0)$ at $x \rightarrow -\infty$ toward the dotted blue circle at $x = 0$. At the bottom of these four panels, we also present the $t$-axis and corresponding values of the selected values of $t$ for $-\frac{1}{2} T_\text{M} < -\frac{\pi}{4} \leq t \leq \frac{\pi}{4} < \frac{1}{2} T_\text{M}$. The trajectories in the upper-part and lower-part of the complex-plane correspond to the positive and negative values of $t$, respectively. We observe that the trajectories shift faster in space around $t = 0$ than around $t = \pm \frac{1}{2} T_\text{M} = \pm \frac{\pi}{\rho}$.

In particular, for $t = n\pi/\rho$, $n \in \mathbb{Z}$, $\tilde{q}_\text{M}$ reduces to a real-valued function, i.e., Im$(\tilde{q}_\text{M}) = 0$ for all $\mu > 0$. Hence, the parameterized curve is a straight line at the real-axis. For $t = 2n \pi/\rho$, $n \in \mathbb{Z}$, this is shown by the horizontal solid red line lying on the real axis moving from a point larger than Re$(\tilde{q}_\text{M}) = 3$ to  Re$(\tilde{q}_\text{M}) = -1$ for $x > 0$. The represented case $t = 0$ is displayed in Figure~\ref{fig4-argkmx-param} while the case $t = \pi/\rho$ is not shown in the figure. Indeed, from~\eqref{KMbreather}, we obtain the following limiting values for $n \in \mathbb{Z}$:
\begin{equation}
\lim\limits_{x \rightarrow 0} q_\text{M}(x,2n\pi/\rho) = 1 + \sqrt{\mu^2 + 4}  \qquad \text{and} \qquad 
\lim\limits_{x \rightarrow 0} q_\text{M}(x,(2n + 1)\pi/\rho) = 1 - \sqrt{\mu^2 + 4}.  
\end{equation}
Additionally, $\lim\limits_{x \rightarrow \pm \infty} q_\text{M}(x,n\pi/\rho) = -1$. Using a similar analysis, vertical straight lines at Re$(\tilde{q}_\text{M}) = -1$ can be obtained by taking the values of $t = (n + \frac{1}{2})\pi/\rho$, for $n \in \mathbb{Z}$. The line direction from the positive and negative regions of Im$(\tilde{q}_\text{M})$ is downward and upward toward $(-1,0)$ for even and odd values of $n \in \mathbb{Z}$, respectively. 
\begin{figure}[hbtp!]
\begin{center}
\subcaptionbox{(a) $\mu = \sqrt{2}$}{\includegraphics[width=0.45\textwidth]{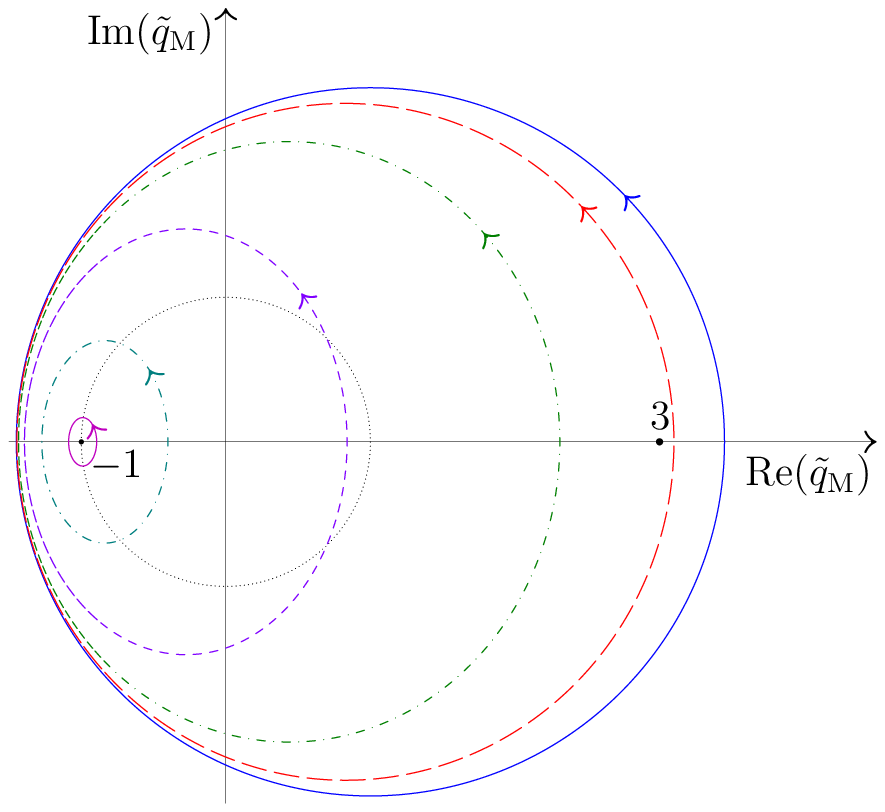}}		\hspace*{1cm}
\subcaptionbox{(b) $\mu = 1$}   	{\includegraphics[width=0.45\textwidth]{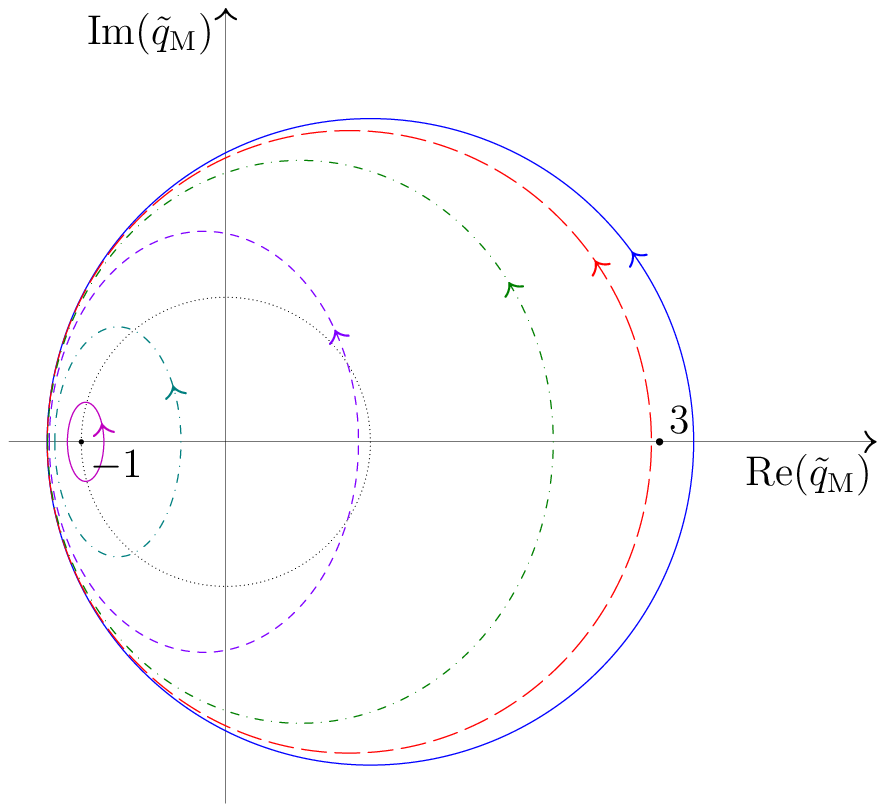}}
\subcaptionbox{(c) $\mu = \frac12$}		{\includegraphics[width=0.45\textwidth]{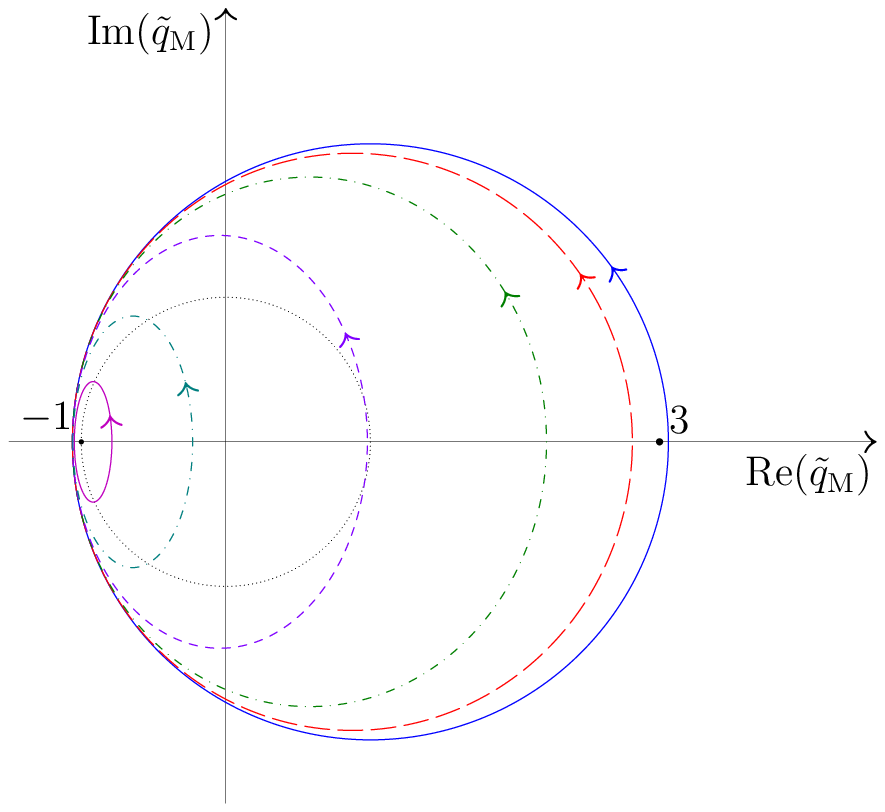}}		\hspace*{1cm}
\subcaptionbox{(d) $\mu = \frac15$}     {\includegraphics[width=0.45\textwidth]{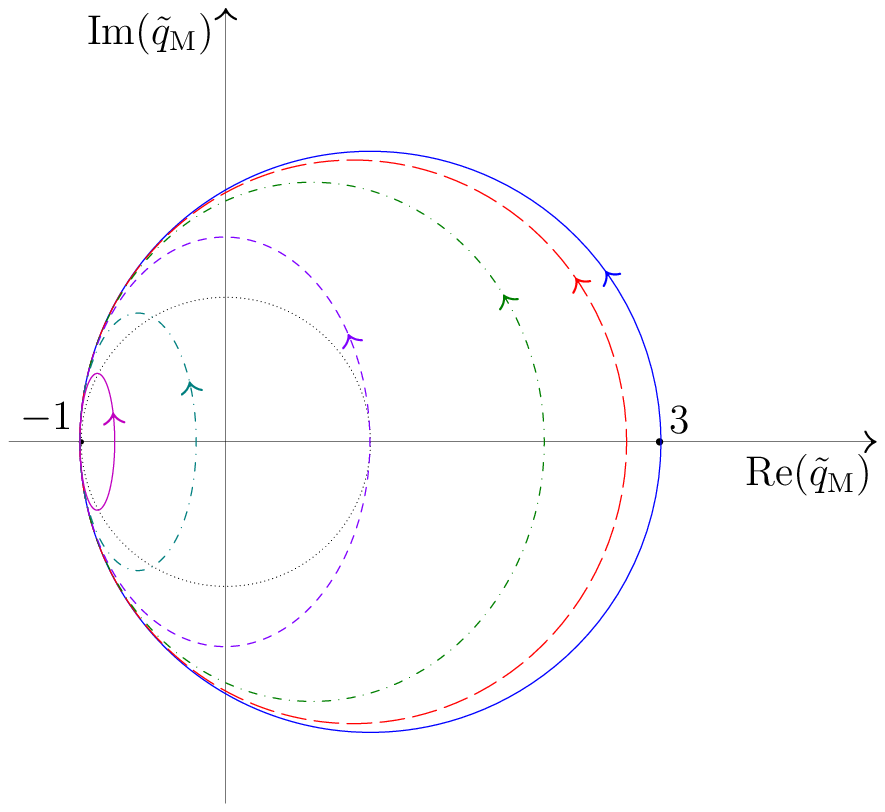}}		\vspace*{0.25cm} 
\subcaptionbox{}{\includegraphics[width=0.9\textwidth]{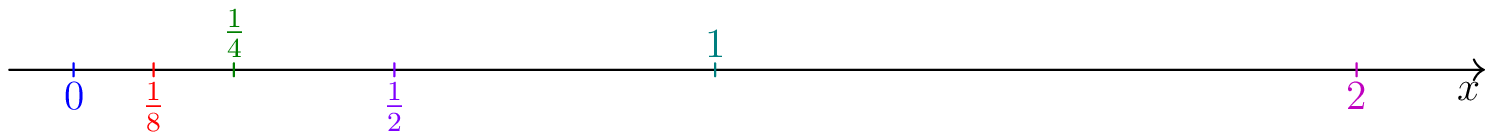}} \vspace*{-0.5cm}
\end{center}
\caption{Parameterization of the non-rapid oscillating complex-valued amplitude of the Kuznetsov-Ma breather $\tilde{q}_\text{M}$ in the temporal variable $t$ ($-\pi/\rho < t < \pi/\rho$) for different values of spatial variable $x$: $x = 0$ (solid blue), $x = 1/8$ (long-dashed red), $x = 1/4$ (dash-dotted green), $x = 1/2$ (dashed purple), $x = 1$ (dash-dotted cyan), and $x = 2$ (solid magenta) and different values of the parameter $\mu$: (a) $\mu = \sqrt{2}$, (b) $\mu = 1$, (c) $\mu = 1/2$, and (d) $\mu = 1/5$.} \label{fig5-argkmt-param}
\end{figure}

\begin{figure}[h!]
\begin{center}
\subcaptionbox{(a) $\nu = 1$}   {\includegraphics[width=0.45\textwidth]{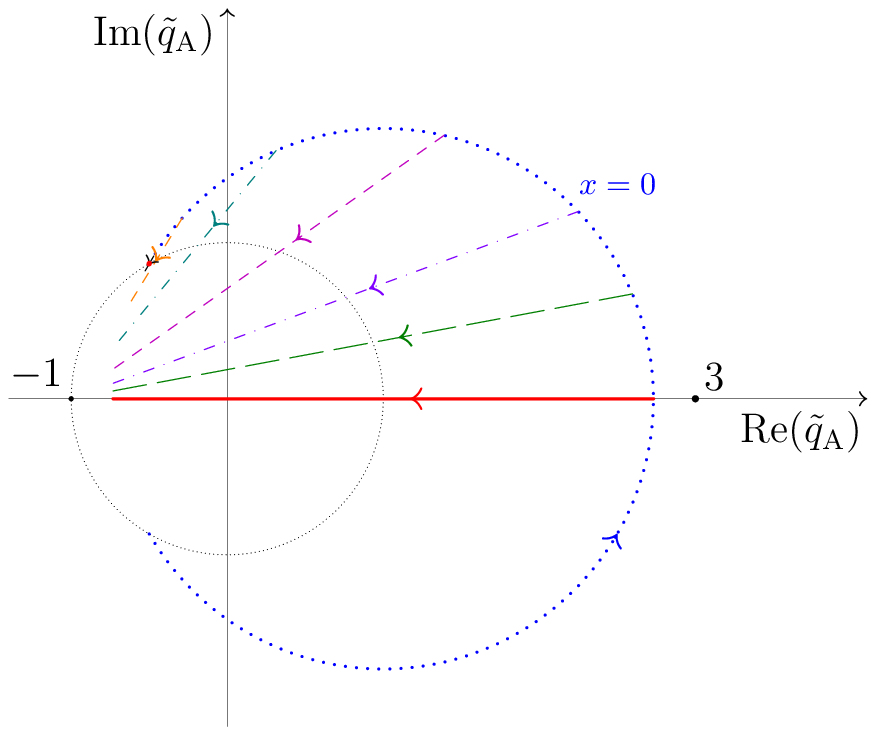}}		\hspace*{1cm}
\subcaptionbox{(b) $\nu = \frac12$} {\includegraphics[width=0.45\textwidth]{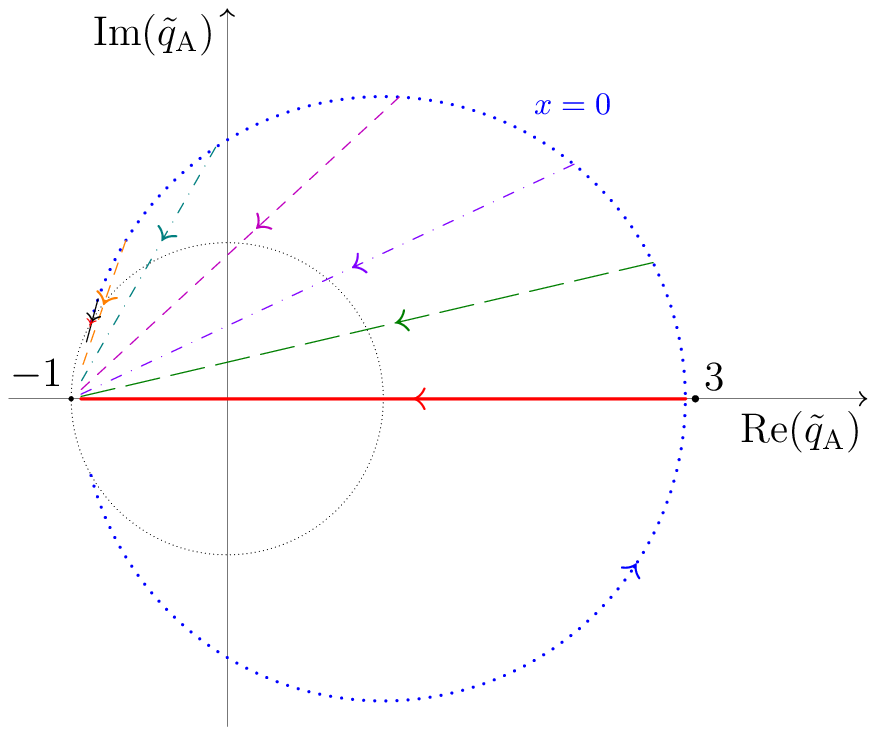}}
\subcaptionbox{(c) $\nu = \frac14$}{\includegraphics[width=0.45\textwidth]{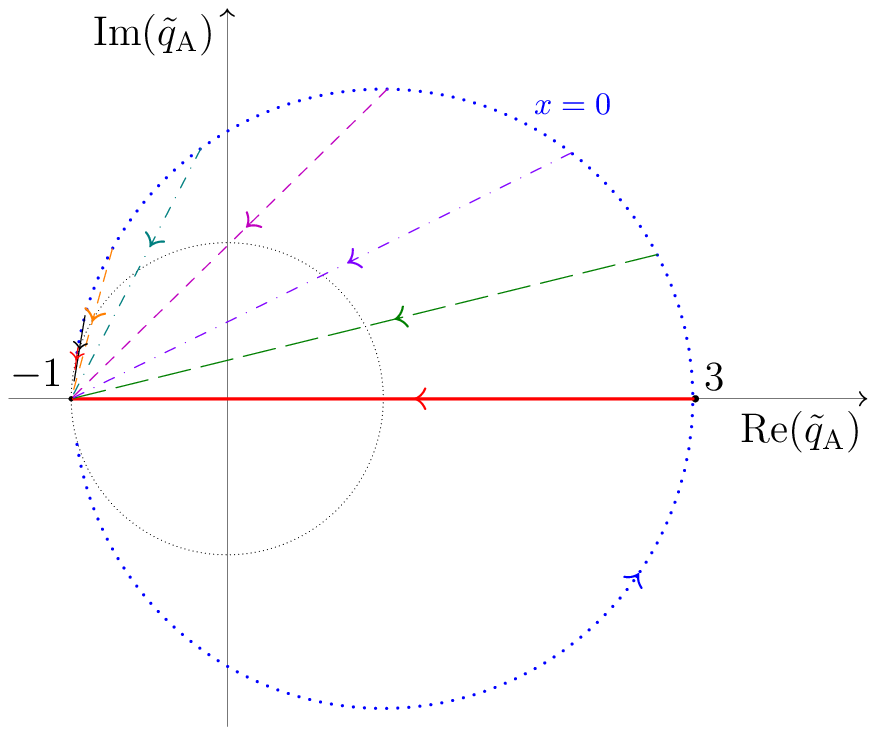}}		\hspace*{1cm}
\subcaptionbox{(d) $\nu = 0$}   {\includegraphics[width=0.45\textwidth]{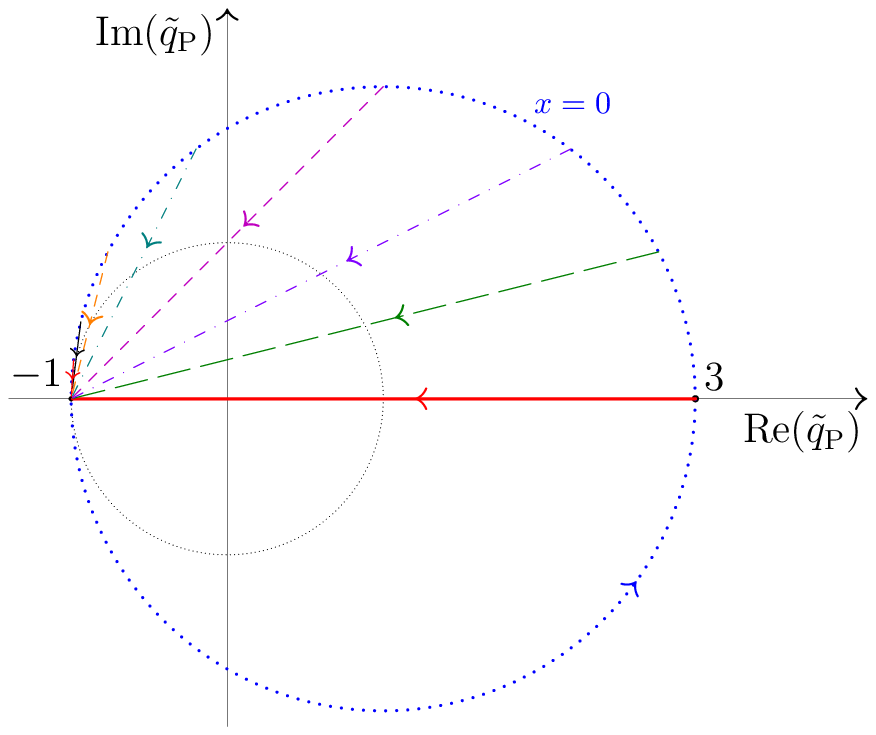}} \vspace*{0.25cm} \\
\subcaptionbox{}{\includegraphics[width=0.9\textwidth]{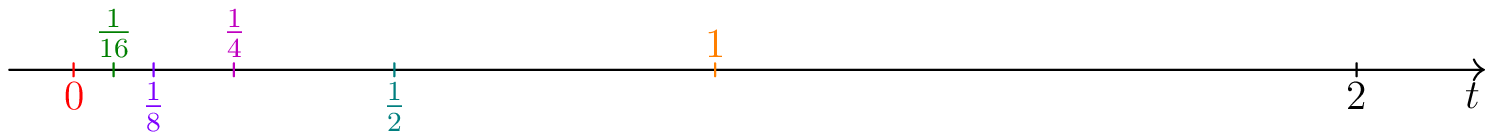}}
\end{center}
\caption{Parameterization of the non-rapid oscillating complex-valued amplitude $\tilde{q}$ in the spatial variable $x$ for different values of temporal variable $t$: $t = 0$ (solid red), $t = 1/16$ (long-dashed green), $t = 1/8$ (dash-dotted purple), $t = 1/4$ (dash magenta), $t = 1/2$ (dash-dotted cyan), $t = 1$ (dashed orange), $t = 2$ (solid black), and $t = 4$ (solid red), and modulation frequencies of the Akhmediev solitons (a) $\nu = 1$ ($0 \leq x \leq \pi$), (b) $\nu = \frac12$ ($0 \leq x \leq 2\pi$), (c) $\nu = \frac14$ ($0 \leq x \leq 4 \pi$), and (d) $\nu = 0$, $x \geq 0$ (Peregrine soliton).} \label{fig6-argwex-param}
\end{figure}

\begin{figure}[h!]
\begin{center}
\subcaptionbox{(a) $\nu = 1$}   {\includegraphics[width=0.45\textwidth]{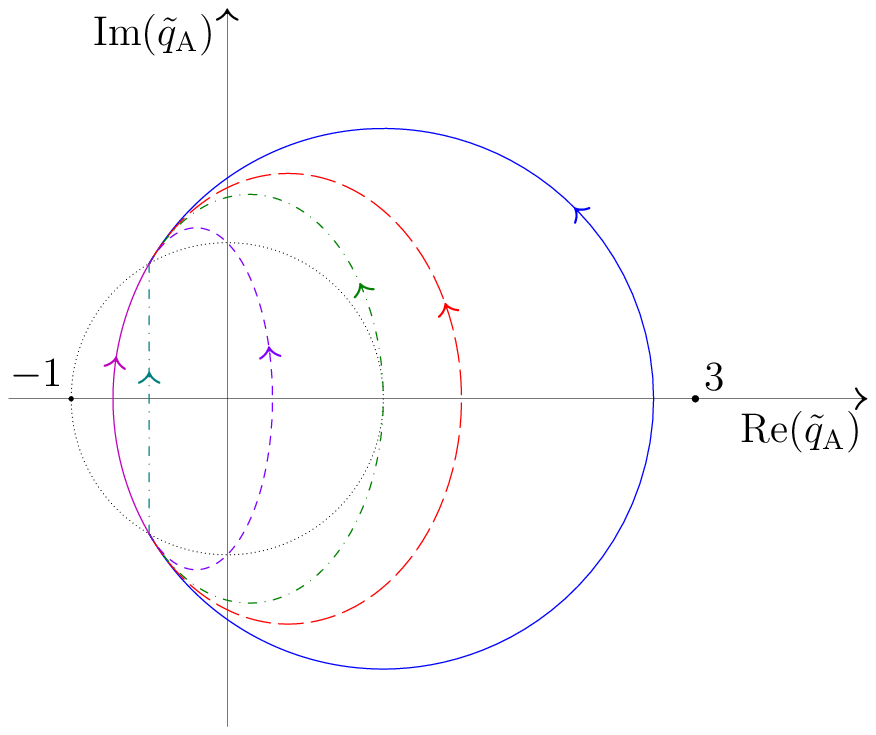}}		\hspace*{1cm}
\subcaptionbox{(b) $\nu = 0.5$} {\includegraphics[width=0.45\textwidth]{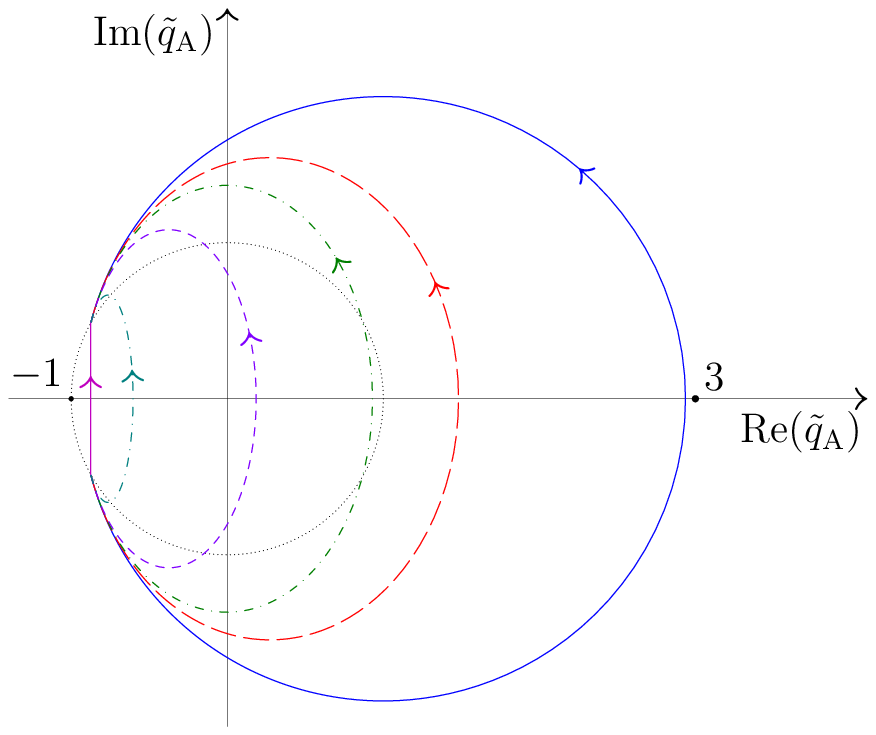}}
\subcaptionbox{(c) $\nu = 0.25$}{\includegraphics[width=0.45\textwidth]{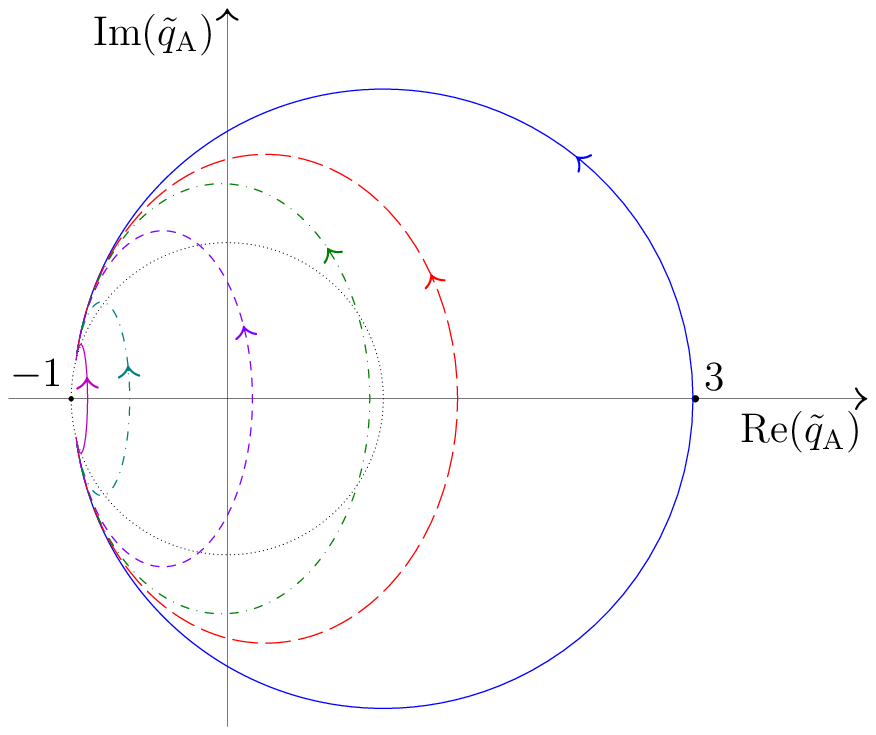}}	\hspace*{1cm}
\subcaptionbox{(d) $\nu = 0$}   {\includegraphics[width=0.45\textwidth]{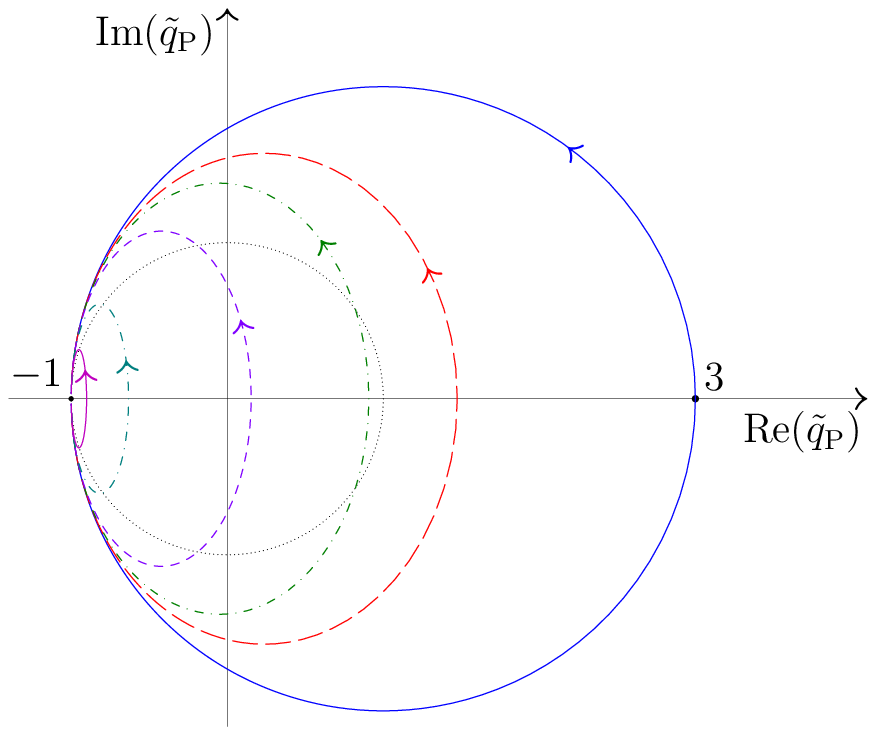}}  \vspace*{0.25cm} \\
\subcaptionbox{}{\includegraphics[width=0.9\textwidth]{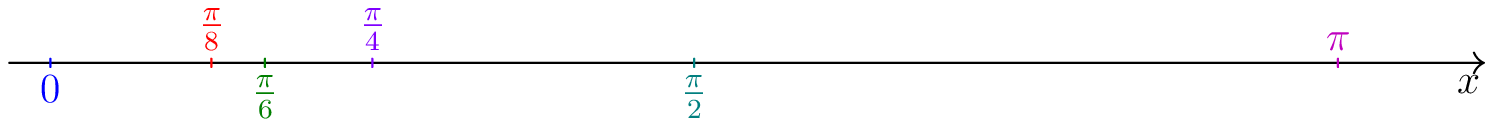}}
\end{center}
\caption{Parameterization of the non-rapid oscillating complex-valued amplitude $\tilde{q}$ in the temporal variable $t$ ($-\infty < t < \infty$) for different values of spatial variable $x$: $x = 0$ (solid blue), $x = \pi/8$ (long-dashed red), $x = \pi/6$ (dash-dotted green), $x = \pi/4$ (dashed purple), $x = \pi/2$ (dash-dotted cyan), and $x = \pi$ (solid magenta), and modulation frequencies of the Akhmediev solitons (a) $\nu = 1$, (b) $\nu = 0.5$, (c) $\nu = 0.25$, and (d) $\nu \rightarrow 0$ (the Peregrine soliton).} \label{fig7-argwet-param}
\end{figure}

\begin{figure}[h!]
\begin{center}
\subcaptionbox{(a) Re$(\tilde{q}_\text{P})$ vs. $x$}{\includegraphics[width=0.45\textwidth]{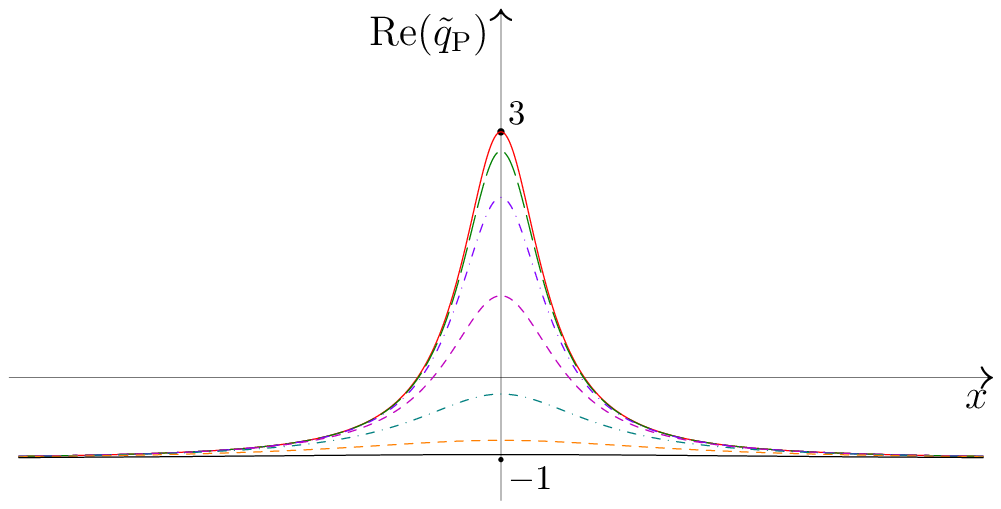}}		\hspace*{1cm}
\subcaptionbox{(b) Im$(\tilde{q}_\text{P})$ vs. $x$}{\includegraphics[width=0.45\textwidth]{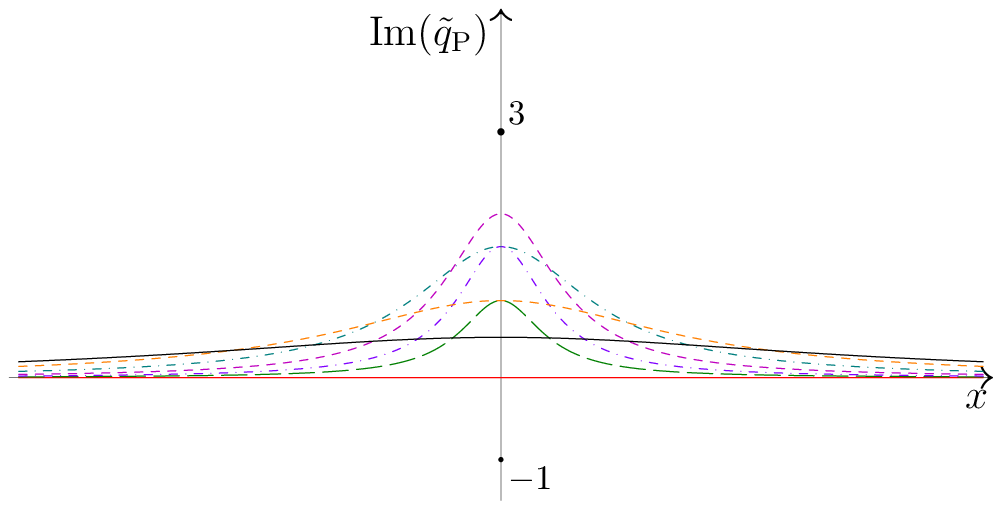}}		
\subcaptionbox{}{\includegraphics[width=0.9\textwidth]{axis-akmi-t}}
\subcaptionbox{(c) Re$(\tilde{q}_\text{P})$ vs. $t$}{\includegraphics[width=0.45\textwidth]{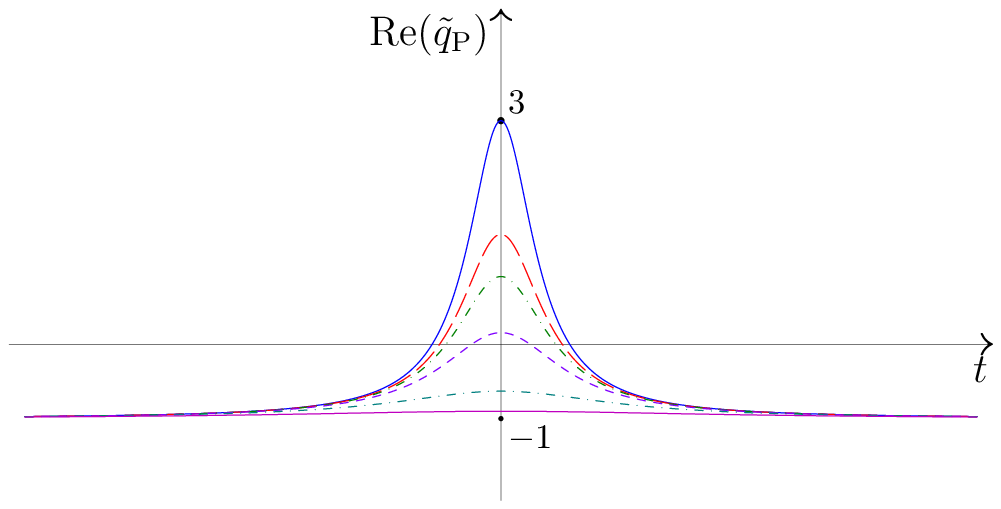}}		\hspace*{1cm}
\subcaptionbox{(d) Im$(\tilde{q}_\text{P})$ vs. $t$}{\includegraphics[width=0.45\textwidth]{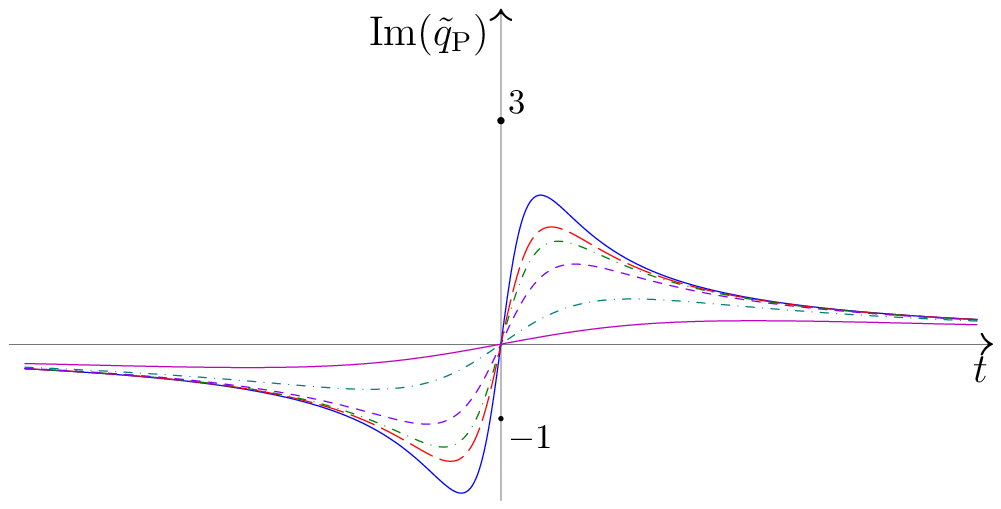}}
\subcaptionbox{}{\includegraphics[width=0.9\textwidth]{axis-akmi-x}}
\end{center}
\caption{Plots of the real and imaginary parts of the non-rapid oscillating complex-valued amplitude for the Peregrine soliton with respect to the spatial and temporal variables, $x$ (upper panels) and $t$ (lower panels), respectively. For upper panels~(a) and~(b), various curves indicates different time: $t = 0$ (solid red), $t = 1/16$ (long-dashed green), $t = 1/8$ (dash-dotted purple), $t = 1/4$ (dash magenta), $t = 1/2$ (dash-dotted cyan), $t = 1$ (dashed orange), and $t = 2$ (solid black). For lower panels~(c) and~(d), different curves indicates different positions: $x = 0$ (solid blue), $x = \pi/8$ (long-dashed red), $x = \pi/6$ (dash-dotted green), $x = \pi/4$ (dashed purple), $x = \pi/2$ (dash-dotted cyan), and $x = \pi$ (solid magenta).} \label{fig8-peratpro}
\end{figure}

Figure~\ref{fig5-argkmt-param} displays the sketch of the non-rapid-oscillating Kuznetsov-Ma breather $\tilde{q}_\text{M}$ in the complex-plane parameterized in the temporal variable $t$ for different values of the spatial variable $x$ and parameter $\mu$. For each case, $t$ is taken for one temporal envelope period, i.e., $-\frac{1}{2}T_\text{M} = -\frac{\pi}{\rho} < t < \frac{\pi}{\rho} = \frac{1}{2} T_\text{M}$. Instead of a set of straight lines, the trajectories form the shape of elliptical curves. For each $x = x_0 \in \mathbb{R}$, the ellipse is centered at $(c(x_0),0)$ with semi-minor axis $a(x_0)$ and semi-major axis $b(x_0)$, where
\begin{align}
a(x_0) &= \frac{\mu \rho \cosh \left(\mu x_0 \right)}{d(x_0)}    		&
b(x_0) &= \frac{\rho \cosh \left(\mu x_0 \right)}{\sqrt{d(x_0)}} \\
c(x_0) &= \frac{2\mu^2}{d(x_0)} - 1									&
d(x_0) &= 2 \cosh\left(2 \mu x_0 \right) + \mu^2 - 2.	
\end{align}

The special case of a circle is obtained for $x_0 = 0$ with the radius $r = \sqrt{\mu^2 + 4}$ centered at $(1,0)$. All curves move in the counterclockwise direction for increasing $t$. For $x > 0$, the larger the values of $x$, the smaller the ellipses become. The situation is the opposite for $x < 0$: smaller values of $x$ (but largely negative in its absolute value sense) correspond to smaller ellipses in the complex plane. Due to its spatial symmetry, only the plots for positive values of $x$ are displayed. The axis below the figure panels shows the selected $x$ values for a better overview of the variable scaling: $x = 0$, $\frac{1}{8}$, $\frac{1}{4}$, $\frac{1}{2}$, $1$, and $x = 2$.

Figure~\ref{fig6-argwex-param} displays the sketch in the complex-plane of the non-rapid-oscillating Akhmediev soliton $\tilde{q}_\text{A}$ [panels~(a)--(c)] and Peregrine soliton $\tilde{q}_\text{P}$ [panel~(d)] parameterized in the spatial variable $x$ for different values of the temporal variable $t$ and parameter $\nu$. We only display the trajectories corresponding to the positive values of $t$, the trajectories for the negative values of $t$ are simply the reflection over the horizontal axis Re$(\tilde{q}_\text{P}) = 0$. The $t$-axis below the panels indicate the chosen values of $t$ displayed in the figure. Similar to the trajectories for the Kuznetsov-Ma breather when they are parameterized in the spatial variable $x$, the trajectories for the Akhmediev soliton parameterized in $x$ are also collections of straight lines shifting in the counterclockwise direction for increasing values of $t$. Different from the previous case, these straight lines are periodic in $x$. The experimental results of deterministic freak wave generation using the spatial NLS equation showed that instead of straight lines, we obtained non-degenerate Wessel curves, suggesting  that there the periodic lines might be perturbed during the downstream evolution~\cite{karjanto2006mathematical,karjanto2010qualitative}. 

For each panel, we only sketch the trajectories for an interval of half the spatial envelope wavelength, i.e., $0 \leq x \leq \frac{1}{2} L_\text{A} = \frac{\pi}{\nu}$. For this limited space interval, the direction of the lines is moving inwardly focused, from the dotted-blue outer circle for $x = 0$ to some values in the left-part of the complex-plane near Re$(\tilde{q}_\text{A}) = -1$. As the value of $x$ progresses, $\frac{1}{2} L_\text{A} = \frac{\pi}{\nu} \leq x \leq L_\text{A} = \frac{2\pi}{\nu}$, the trajectories bounce back toward the initial points by following the identical paths. They then travel in the same manner periodically as $x \rightarrow \pm \infty$. For a decreasing value of the parameter $\nu$, the endpoint of these lines tends to focus around the region near $(-1,0)$, as we can observe in Figures~\ref{fig6-argwex-param}(a)--\ref{fig6-argwex-param}(c). For the Peregrine soliton, the trajectories are not periodic as $L_\text{A} \rightarrow \infty$ and they tend to $(-1,0)$ for $x \rightarrow \pm \infty$, as can be seen in Figure~\ref{fig6-argwex-param}(d).

Figure~\ref{fig7-argwet-param} displays the sketch of the non-rapid-oscillating part of the Akhmediev soliton $\tilde{q}_\text{A}$ [panels~(a)--(c)] and Peregrine soliton $\tilde{q}_\text{P}$ [panel~(d)] in the complex-plane parameterized in the temporal variable $t$ for different values of the spatial variable $x$ and parameter $\nu$. The values of $t$ run from $t \to -\infty$ to $t \to +\infty$, and we only sketch the positive values of $x$. The plots for the negative values of $x$ are identical and are not shown due to the symmetry property of the soliton. The $x$-axis below the panels shows the selected values of $x$ ranging from $x = 0$ to $x = \pi$. For $\tilde{q}_\text{A}$, the trajectories are composed of circular sectors, elliptical sectors, and straight lines instead of closed curves like circles or ellipses. Since this soliton is a nonlinear extension of the modulational instability, the trajectories for each value of the parameter $\nu$, $0 < \nu < 2$, are the corresponding homoclinic orbit for an unstable mode, and the presence of a phase shift prevents closed-path trajectories~\cite{ablowitz1990homoclinic,akhmediev1997solitons,calini2002homoclinic,kimmoun2016modulation}. 

The circular sectors are attained for $x = 0$ and the straight lines occur at $x = \left(n + \frac{1}{2}\right) \frac{\pi}{\nu}$, $n \in \mathbb{Z}$. Trajectories at other locations yield the elliptical sectors. The initial and final points are not identical and this indicates a phase shift in the soliton. Let $\phi_{+\infty}$ and $\phi_{-\infty}$ be the phases for $x \rightarrow \pm \infty$, respectively. Let also $\Delta \phi = \phi_{+\infty} - \phi_{-\infty}$ be the difference between the phase at $x = +\infty$ and $x = -\infty$, then we have the following phase relationships:
\begin{equation}
\tan \phi_{\pm \infty} = \pm \frac{\sigma}{\nu^2 - 1} \qquad \text{and} \qquad \Delta \phi = 2 \arctan \left(\frac{\sigma}{\nu^2 - 1} \right).
\end{equation}

For the Peregrine soliton, the trajectories of time parameterization in the complex-plane are either a circle (for $x = 0$) or ellipses (for other values of $x \neq 0$). The circle is centered at $(1,0)$ with radius $r = 2$. Let $x = x_0 \in \mathbb{R}$ be the position for the Peregrine soliton, then the ellipse has the length of semi-minor axis $a(x_0)$, the length of semi-major axis $b(x_0)$, and is centered at $(c(x_0),0)$, where
\begin{equation}
a(x_0) = \frac{2}{1 + 4x_0^2}, \qquad \qquad b(x_0) = \frac{2}{\sqrt{1 + 4x_0^2}}, \qquad \text{and} \qquad c(x_0) = a(x_0) - 1.
\end{equation}

Figure~\ref{fig8-peratpro} should be viewed in connection to Figures~\ref{fig6-argwex-param}(d) and \ref{fig7-argwet-param}(d). It displays the plots of the real and imaginary parts of the non-rapid-oscillating complex-valued amplitude for the Peregrine soliton $\tilde{q}_\text{P}$ with respect to $x$ and $t$, which are presented in the upper and lower panels, respectively. For the former, different curves correspond to selected values of time $t \in \left\{0, \frac{1}{16}, \frac{1}{8}, \frac{1}{4}, \frac{1}{2}, 1, 2 \right\}$. For the latter, different curves correspond to selected values of position $x \in \left\{0, \frac{\pi}{8}, \frac{\pi}{6}, \frac{\pi}{4}, \frac{\pi}{2}, \pi \right\}$. The phase difference in the time parameterization of $\tilde{q}_\text{P}$ is discernible from the behavior of Im$(\tilde{q}_\text{P})$ as $t \rightarrow \pm \infty$. While $\lim\limits_{t \rightarrow \pm \infty} \text{Re}(\tilde{q}_\text{P}) = -1$, the quantity for $\lim\limits_{t \rightarrow \pm \infty} \text{Im}(\tilde{q}_\text{P})$ takes positive and negative values, respectively.

\section{Conclusion} 		\label{conclusion}

We have considered the exact analytical breather solutions of the focusing NLS equation, where the wave envelopes at infinity have a nonzero but constant background. These solutions have been adopted as weakly nonlinear prototypes for freak wave events in dispersive media due to their fine agreement with various experimental results. We have provided not only a brief historical review of the breathers but also covered some recent progress in the field of rogue wave modeling in the context of the NLS equation. 

In particular, we have discussed the Peregrine soliton as a limiting case of the Kuznetsov-Ma breather and Akhmediev soliton. We have verified rigorously using the $\epsilon$-$\delta$ argument that as each of the parameter values from these two breathers is approaching zero, they reduce to the Peregrine soliton. We have also presented this limiting behavior visually by depicting the contour plots of the breather amplitude modulus for selected parameter values. We displayed the parameterization plots of the non-rapid-oscillating complex-valued breather amplitudes both spatially and temporally. 

The trajectories for the spatial parameterization in the complex-plane exhibit a set of straight lines for all the breathers. From $x \to -\infty$ to $x \to +\infty$, the paths are passed twice for the Kuznetsov-Ma breather and are elapsed many times infinitely for the Akhmediev soliton due to its spatial periodic characteristics. The trajectories in the complex plane for the parameterization in the temporal variable of the Kuznetsov-Ma breather and Peregrine soliton feature a periodic circle and a set of periodic ellipses due to its temporal symmetry. For the Akhmediev soliton, on the other hand, the path does not only turn into circle and ellipse sectors but also becomes straight lines as it travels from $t \to -\infty$ to $t \to +\infty$, featuring homoclinic orbits with a phase shift. 

\section*{Conflict of Interest Statement}
The authors declare that the research was conducted in the absence of any commercial or financial relationships that could be construed as a potential conflict of interest.

\section*{Author Contributions}
The author has contributed to completing this article.

\section*{Funding}
This research does not receive any funding.

\section*{Acknowledgments}
The author wishes to thank Bertrand Kibler, Amin Chabchoub, and Heremba Bailung for the invitation to contribute to the article collection ``Peregrine Soliton and Breathers in Wave Physics: Achievements and Perspectives''. The author also acknowledges E. (Brenny) van Groesen, Mark Ablowitz, Constance Schober, Frederic Dias, Roger Grimshaw, Panayotis Kevrekidis, Boris Malomed, Evgenii Kuznetsov, Nail Akhmediev, Alfred Osborne, Miguel Onorato, Gert Klopman, Rene Huijsmans, Andonowati, Stephan van Gils, Guido Schneider, Anthony Roberts, Shanti Toenger, Omar Kirikchi, Ardhasena Sopaheluwakan, Hadi Susanto, Alexander Iskandar, Agung Trisetyarso, and Defrianto Pratama for fruitful discussion.

\section*{Supplemental Data}
There is no supplemental data for this article.

\section*{Data Availability Statement}
There is no available data for this article.

\bibliographystyle{frontiersinHLTH_FPHY} 
\bibliography{2006fro}
\end{document}